\documentclass[apjs,iop]{emulateapj}
\usepackage{color}
\usepackage{hyperref}
\hypersetup{
pdftitle={LAMOST DR1: stellar parameters and chemical abundances with
SP\_Ace},
pdfauthor={Corrado Boeche},
colorlinks=true,
linkcolor=blue,
citecolor=blue,
urlcolor=blue
}

\newcommand\temp{$T_{\fontsize{6}{6}\selectfont \mbox{eff}}$}\normalfont
\newcommand\logg{$\log g$}

\newcommand\met{[M/H]}
\newcommand\aFe{[$\alpha$/Fe]}
\newcommand\Space{SP\_Ace}
\newcommand\kmsec{km s$^{-1}$}

\newcommand\SNspace{S/N$_{\rm SPAce}$}

\begin{document}

\title{LAMOST DR1: stellar parameters and chemical abundances with \Space}

\author{C. Boeche}
\affil{Astronomisches Rechen-Institut, Zentrum f\"ur Astronomie der
Universit\"at Heidelberg, M\"onchhofstr. 12-14, 69120 Heidelberg, Germany}
\affil{INAF, Padova Observatory, Vicolo dell'Osservatorio
5, 35122 Padova, Italy}
\email{Email: corrado@ari.uni-heidelberg.de, msmith@shao.ac.cn}

\author{M. C. Smith}
\affil{Key Laboratory for Research in Galaxies and Cosmology, Shanghai
Astronomical Observatory, Chinese Academy of Sciences, 80 Nandan Road,
Shanghai 200030, China}

\author{E. K. Grebel}
\affil{
Astronomisches Rechen-Institut, Zentrum f\"ur Astronomie der
Universit\"at Heidelberg, M\"onchhofstr. 12-14, 69120 Heidelberg, Germany}

\author{J. Zhong}
\affil{Key Laboratory for Research in Galaxies and Cosmology, Shanghai
Astronomical Observatory, Chinese Academy of Sciences, 80 Nandan Road,
Shanghai 200030, China}

\author{J.L. Hou}
\affil{Key Laboratory for Research in Galaxies and Cosmology, Shanghai
Astronomical Observatory, Chinese Academy of Sciences, 80 Nandan Road,
Shanghai 200030, China}
\affil{University of Chinese Academy of Sciences, Beijing 100049, China}

\author{L. Chen}
\affil{Key Laboratory for Research in Galaxies and Cosmology, Shanghai
Astronomical Observatory, Chinese Academy of Sciences, 80 Nandan Road,
Shanghai 200030, China}
\affil{University of Chinese Academy of Sciences, Beijing 100049, China}

\and

\author{D. Stello}
\affil{School of Physics, University of New South Wales, NSW 2052,
Australia}
\affil{Sydney Institute for Astronomy (SIfA), School of Physics, University
of Sydney, NSW 2006, Australia}
\affil{Stellar Astrophysics Centre, Department of Physics and Astronomy,
Aarhus University, Ny Munkegade 120, DK-8000 Aarhus C, Denmark}

\begin{abstract}
We present a new analysis of the LAMOST DR1 survey spectral database performed with
the code \Space, which provides the derived stellar parameters \temp,
\logg, [Fe/H], and [$\alpha$/Fe] for 1\,097\,231 stellar objects. We
tested the reliability of our results by comparing them to reference
results from high spectral resolution surveys. The expected errors can
be summarized as $\sim$120~K in 
\temp, $\sim$0.2 in \logg, $\sim$0.15~dex in [Fe/H], and
$\sim$0.1~dex in [$\alpha$/Fe] for spectra with S/N$>$40, with
some differences between dwarf and giant stars. \Space\ provides error
estimations consistent with the discrepancies observed between derived
and reference parameters. Some systematic errors are identified and
discussed. The resulting catalog is publicly available at the LAMOST
and CDS websites.
\end{abstract}


\section{Introduction}

During the last decades the formation and evolution of the Milky Way
(MW) has become a question of major importance in modern astrophysics. The
proximity of our Galaxy permits individual star-by-star investigations that
would not be possible for external galaxies. This opportunity has been
grasped by many research groups who have planned and run spectroscopic
surveys of the MW at high- and low-spectral resolution (e.g., the Apache Point
Observatory Galactic Evolution Experiment, APOGEE, \citealp[Allende Prieto
et al.][]{apogee}; the Galactic Archaeology with
HERMES (GALAH) Survey, \citealp[Zucker et al.][]{zucker}; the Gaia-ESO Public
Spectroscopic Survey, \citealp[Gilmore et al.][]{gilmore}; 
the Sloan Extension for Galactic
Understanding and Exploration, SEGUE, \citealp[Yanny et
al.,][]{yanny}; the RAdial Velocity Experiment, RAVE, \citealp[Steinmetz
at al.][]{rave}; the 4-Metre
multi-Object Spectroscopic Telescope, 4MOST, \citealp[de Jong et
al.][]{dejong}; the WHT Enhanced Area Velocity Explorer, WEAVE, 
\citealp[Dalton et al.][]{dalton}).\\
While high-resolution spectroscopy can provide higher accuracy in stellar
parameters and chemical abundances for relatively small 
samples of stars that are relatively bright (e.g. $10^6$ stars brighter than
V=14 for GALAH), low-resolution spectroscopy is better suited to
collecting spectra of stars of fainter magnitude, thus securing a much
larger sample ($25\cdot10^6$ stars down to V=20 for 4MOST).\\
The Large Sky Area Multi-Object Fiber Spectroscopic 
Telescope (LAMOST, \citealp[Cui et al.][]{cui2012}) is a telescope with
effective aperture of 4m, which is used to conduct Galactic and
extra-galactic spectroscopic surveys at spectral resolution of R
$\sim$ 2000.
The LAMOST Experiment for Galactic Understanding and Exploration (LEGUE,
Deng et al., \citealp{deng2012}) is an on-going Galactic survey that, with a
present sample of more than 5 million stellar spectra of the MW, is
the largest spectroscopic survey available to the astronomical
community. The first data release (DR1, Luo et al. \citealp{luo})
includes 2\,204\,696 spectra, of which 1\,944\,329 are 
spectra of stars in the MW (corresponding to around 1.6M unique
stars). Public data releases have followed on an annual
basis\footnote{http://www.lamost.org}.\\

In our current paper we present the stellar parameters and chemical abundances 
obtained by applying the code \Space\ (Boeche \& Grebel, \citealp{boeche}, see
also Section~\ref{sec:space}) to the LAMOST DR1 spectra. Stellar
parameters, such as \temp, \logg, and \met\ have already been derived,
most notably by the official LAMOST pipeline, LASP (Luo et al.,
\citealp{luo}). This pipeline initially compares the observed spectra
to a library of synthetic spectra derived from a Kurucz/ATLAS9 grid in
order to obtain a first, coarse, estimate of the parameters. After
this the {\it ULySS} method (Koleva et al. \citealp{koleva}, 
Wu et al. \citealp{wu}), using the ELODIE spectral library (Prugniel et
al. \citealp{prugniel}) for the template
spectra, is applied.  In addition to LASP, other groups have carried out   
their own analyses. Ho et al. (\citealp{ho},\citealp{ho2}) have applied a method based on the
``The Cannon" (Ness et al., \citealp{ness}), which estimates
parameters by training a model on existing data sets (in this instance
from the APOGEE survey).
From this analysis they are able to deriving parameters for giant
stars, including [$\alpha$/Fe], individual [C/M] and [N/M]
abundances and, from these latter two abundances, masses and ages.
Xiang et al. \cite{xiang} present another analysis, like LASP
also matching to template spectra (in this case observed libraries from
MILES and ELODIE). A recent update to this pipeline now delivers
alpha-element abundances and, by applying a machine learning algorithm
to the giant stars, [C/N] and [N/H] abundances (Xiang et al. \citealp{xiang2017}).
By using \Space\ we derive the stellar parameters \temp, \logg, [Fe/H], and
the alpha abundances [$\alpha$/Fe] for both dwarf and giant stars.

\begin{figure*}
\centering
\includegraphics[bb=75 295 500 442]{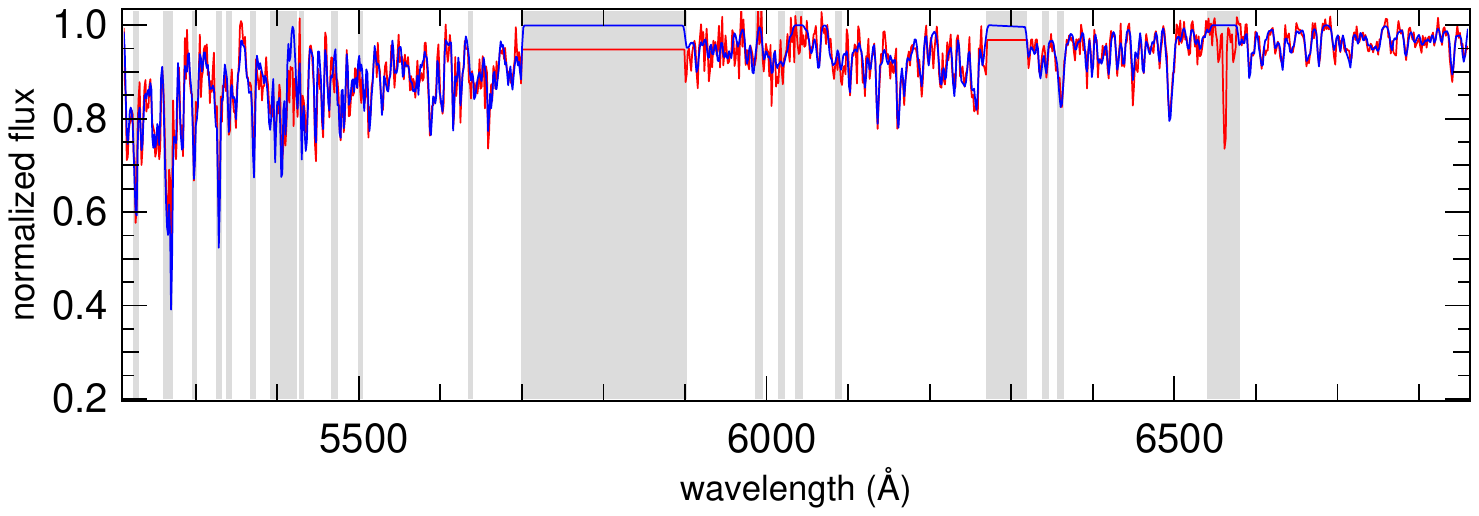}
\caption{LAMOST spectrum spec-55932-GAC\_061N46\_V1\_sp03-061 (in red) and the best
matching model found by \Space\ (in blue) with derived parameters of \temp=4642~K,
\logg=2.60, \met=0.24~dex. The observed spectrum has S/N=44 and it has
been continuum re-normalized with the internal \Space\ procedure. Shaded areas are the
rejected wavelength intervals.}
\label{spectrum_fitting}
\end{figure*}

\section{Data}\label{sec:data}

We employ the LAMOST spectra from the first data release (DR1, Luo et al.,
\citealp{luo}) using the latest internal DR3 reduction.
Out of the 2\,204\,969 spectra belonging to the DR1 catalog, 
1\,944\,329 were classified as stars, the rest as galaxies, quasars, or
other non-stellar objects.
These spectra are radial velocity corrected for the Earth motion
(only), flux calibrated, and are the result of joining
the two spectra obtained by the blue and red arms of the
spectrographs (Luo et al., \citealp{luo}). The spectra are
re-binned to a constant velocity dispersion so that the pixel interval is
constant in $\log \lambda$.
The calibration is in the vacuum wavelength.
To process these spectra with \Space\ and derive
stellar parameters and chemical abundances we must re-shape the spectra by:
i) converting the dispersion from $\log \lambda$
to $\lambda$ (angstrom), ii) converting the wavelength from vacuum to air, and
iii) normalizing the flux.
We used the IRAF\footnote{Image Reduction and Analysis Facility,
http://iraf.noao.edu.} task {\it disptrans} to convert the wavelength calibration
from vacuum to air and from $\log \lambda$ to $\lambda$. This conversion
renders a spectral dispersion that is not constant. To flux normalize
the spectra we used the IRAF task {\it continuum} with the settings\\

{\it functio=spline3 order=3 low\_rej=1 high\_rej=3 niterat=5}.\\

Out of the 2\,204\,969 DR1 spectra,
we processed 2\,052\,662 spectra; 152\,300 were not present in the internal 
DR3 catalog, mostly objects with low ($<20$) signal-to-noise ratio
(S/N), while 7 spectra failed the IRAF tasks described above.
We ran all of the spectra through \Space, including the ones classified as
non-stellar objects to cover possible mis-classifications.
Because \Space\ was designed for stellar spectra, we expect no
convergence (and therefore null values to be output) for spectra
of objects that are not stars (this is discussed in Section~\ref{sec:class}).

\section{The \Space\ code}\label{sec:space}

The \Space\ software (Boeche \& Grebel, \citealp{boeche})  derives stellar 
parameters and chemical abundances of FGK stars from the analysis of their spectra 
in the wavelength windows 5212-6960\AA\ and 8400-8920\AA. The
software uses a novel method to derive these parameters. Many 
pipelines rely on equivalent width ($EW$) measurements (among others, Fast Automatic 
Moog Analysis, FAMA, \citealp[Magrini et
al.][]{magrini}; GALA, \citealp[Mucciarelli et al.][]{mucciarelli}) or on 
libraries of synthetic spectra (among others, the
MATrix Inversion for Spectral SynthEsis, MATISSE,
\citealp[Recio-Blanco et al.][]{recio-blanco}; FERRE, \citealp[Allende
Prieto et al.][]{allendeprieto}), while others rely on training sets of
standard stars (The Cannon, Ness et al. \citealp{ness}; ULySS,
\citealp[Koleva et al.][]{koleva}; LASP, Luo et al. \citealp{luo}).
Unlike these, \Space\ relies on a library of general curves-of-growth
(GCOG). The GCOG is a function that describes the $EW$ of an absorption
line as a function of the stellar parameters
\temp, \logg, and [El/H], where ``El" is the element the line belongs to. 
The GCOG is, therefore, the generalization of the
classical curve-of-growth extended to the parameters \temp\ and \logg.
Given the stellar parameters \temp, \logg, and
the abundances [El/H], \Space\ takes the GCOG of the lines stored in
the GCOG library for the wavelength interval under consideration and
computes the expected $EW$s. Then, assuming a Voigt line profile,
\Space\ constructs a model spectrum by subtracting these line
profiles (with the just-computed $EW$s) from a continuum normalized
to 1. This is described in more detail in section~7.1 of 
Boeche \& Grebel \citealp{boeche} (see also
Fig.~\ref{spectrum_fitting} of the current paper). In this
way  \Space\ constructs many model spectra of different
stellar parameters and searches for the one that minimizes the $\chi^2$ between
the observed and the model spectrum. The $\chi^2$ is minimized following the 
Levenberg-Marquadt method, details of which are given in section~7.3 of
Boeche \& Grebel \cite{boeche}. This method of comparing model spectra and
observed spectra puts \Space\ in the ``global fitting
methods" category. We remind the reader that as input \Space\
takes spectra that have been wavelength calibrated, radial velocity
corrected to the rest frame, and continuum normalized. However, \Space\ can
apply changes in radial velocity (RV) to the model
spectrum (see section 7.3 of the Boeche \& Grebel \citealp{boeche}
paper). This feature was implemented in the code because experience
showed that \Space\ could detect small wavelength shifts in observed
spectra, even when these were previously RV corrected to the rest frame.
Since this can badly affect the $\chi^2$ analysis,
we allowed \Space\ to perform small shifts (no larger than $\sim$1~FWHM) 
to the spectrum model. This was supposed to be a mere internal setting performed in
order to better match the observed spectrum and increase the accuracy
of the derived stellar parameters.
In fact, in the case of high resolution spectra, a shift of 1 FWHM in wavelength
corresponds to a small shift in RV (at R$\sim$20\,000 this corresponds to
$\sim$15 \kmsec). The case of low resolution spectra (like LAMOST) is
different, because a 1~FWHM shift corresponds to $\sim$200 \kmsec\ in RV. Because the
LAMOST spectra are corrected for the Earth's motion only, these
internal RV corrections correspond to the heliocentric RVs of the LAMOST stars.
From now on, we
refer to these shifts as the \Space\ RV corrections. However, \Space\ was not
designed to measure RVs and has never been tested for this purpose,
therefore will not use them as such. Official LAMOST RVs 
have been determined by the LAMOST 1D pipeline (see Luo et al
\citealp{luo}).\\
Similarly, \Space\ applies a continuum re-normalization to the (already
normalized) observed spectra as an internal trim of the
continuum to improve the $\chi^2$ analysis.
This internal setting was implemented because the normalization of spectra 
(done with IRAF or similar tools) are often not
optimal in the case of wide absorption lines and/or high metallicity.
This is due to the difficulty in distinguishing the continuum level from a
pseudo-continuum generated by the wide blends of lines or wings of
wide strong lines. 
However, in order to deal with certain idiosyncrasies
of the LAMOST spectra, we have changed some parts of the code, as
explained in the following section.

\subsection{The LAMOST version of \Space}\label{sec:SPAce-LAMOST}

To fully exploit the information carried by the LAMOST spectra
we need to consider the largest wavelength range possible. This means that
both the blue and red parts of the spectra must be used for the analysis.
However, the blue and red parts differ from each other as follows:
\begin{itemize}
\item Spectral dispersions: The LAMOST spectra have been rebinned
to a wavelength width that is constant in $\log \lambda$. This means that
the binning in wavelength (employed by \Space) varies along the whole
spectrum.
\item S/N: The intensity of the blue and the red parts of a spectrum
change not only as a function of the temperature of the stars, but also because the
efficiency of the two spectrograph arms differs. Therefore the S/N of
these two parts is different. Furthermore the S/N changes inside each
of the two parts because the central part of the spectral orders
receives more light than the borders (blaze function).
\item Spectral resolutions: The instrumental full-width-half-maximum (FWHM)
of the blue and red arm of the spectroscope differ, with the former being
smaller than the latter.
\item Wavelength calibration: There are differences in the wavelength
calibration between the blue and red parts of the spectra, since
they are done independently. This can lead to small differences in
RV.
\end{itemize}

The most recent public version of \Space\ is able to
handle spectra with a dispersion and S/N varying along the spectrum.
However, for the analysis of a given spectrum, the public version of \Space\ 
assumes one single FWHM and radial velocity correction.
This makes the public version of \Space\ unsuitable for a full
exploitation of the LAMOST spectra.
To overcome this limitation, we have
made a custom version of \Space\ capable of analyzing the blue and red parts
of the spectra separately. This version estimates an individual radial
velocity correction and
instrumental FWHM for each of the two parts, while simultaneously
searching for a single set of stellar parameters and chemical
abundances with a unique $\chi^2$ analysis. 
Unlike the public version of \Space, this version also
makes use of the H$\alpha$ absorption line because it improves
the parameter estimation of the LAMOST spectra for dwarf stars.
For giant stars deviations from local thermodynamic equilibrium
(non-LTE) can affect H$\alpha$ significantly, but for dwarf stars
(i.e., stars with high gravities, \logg$>$4) the LTE assumptions are
good enough to predict a reliable strength for this line.
Therefore, the LAMOST version of \Space\ uses this line by assigning
different weights to the pixels associated with the line\footnote{
We classify pixels as belonging to the H$\alpha$ line if they are
within 3$\sigma$ of the centre of the line, where we assume that the
line profile is approximately Gaussian (i.e. 1 FWHM =
2.35$\sigma$).} as follows,
\begin{eqnarray}
weight= &1 & \mbox{ for }\log g\geq4\\\nonumber
weight= & 1-(4-\log g) & \mbox{ for }3<\log g<4\\\nonumber
weight= & 0 & \mbox{ for }\log g\leq 3.\\ \nonumber
\end{eqnarray}
This provides a significant improvement in \logg\ 
for dwarf stars, while leaving the analysis of giant stars unaffected. 
For this work only, the H$\alpha$ GCOG employed has been obtained
with the same procedure followed for the other lines, but
we used the spectrum synthesis code SPECTRUM (Gray \& Corbally, \citealp{gray}) 
instead of MOOG (Sneden \citealp{sneden}). This was done because, with the
same atomic parameters and stellar atmospheres employed as for the
official \Space\ GCOG library, the strength of the H$\alpha$ line
synthesized with SPECTRUM is bigger than the one in MOOG and it provides
a better match to that of the Sun and Procyon. For this line only, we
adopted the approximated Voigt profile by Bruce et al. \cite{bruce},
\begin{equation}\label{eq_voigt}
Voigt(x)=EW\cdot [rL(x)+(1-r)G(x)],
\end{equation}
where $L$ and $G$ are the Lorentzian and Gaussian functions and $EW$ is the
equivalent width expressed in \AA. While the FWHM of the Gaussian is equal
to the spectral resolution FWHM (one of the variables that \Space\ actively looks
for), the $\sigma$ of the Lorentzian profile depends on the EW
according to the following relation,
\begin{equation}
\sigma_{L}=EW\cdot\Big(1-\exp\big(-(2EW)^2\big)\Big)
\end{equation}
where $\sigma_{L}$ is expressed in \AA, 
and the $r$ parameter in equation \ref{eq_voigt} is defined as,
\begin{equation}
r=\frac{1}{\exp\Big(\frac{1\AA}{2EW+0.001}\Big)}.
\end{equation}

These are empirical relations that have been chosen
to provide an optimal match to the observed H$\alpha$ line profile.

Although the LAMOST catalog provides an estimate of the error on the
flux of a spectrum (inverse variance), it has been noted that this
overestimates the true uncertainty (e.g., Ho et al. \citealp{ho},
Xiang et al. \citealp{xiang}).
If we use these values in our fitting this will result in unreliable
parameter uncertainties and so we have chosen to estimate the error
ourselves. The uncertainty on the i-th pixel is taken as the standard
deviation of the residuals computed over an interval with 25 pixels
of half-width and centered on the i-th pixel (after the
pixels deviating more than 3$\sigma$ were rejected).
This approach ensures that the best-fit reduced $\chi^2$ is
approximately one for most spectra and hence the parameter
uncertainties should be reliable.

With this version of \Space\ we processed the 2\,052\,662
LAMOST DR1 stellar spectra, along with more than 50\,000 spectra from
later data releases. This latter sample consists of spectra of stars with
existing stellar parameters in the literature, chosen so that we can
validate our results (see Section~\ref{sec:validation}).
We limited the analysis to the wavelength intervals
5212--5700\AA, 5900--6270\AA, and 6320--6860\AA. The first neglected
interval (5700--5900\AA) avoids the overlapping region of the blue and red
arm, while the second one (6270--6320\AA) avoids the region affected by
telluric lines.  The processing was carried out with
the \Space\ options ``ABD\_loop" and ``alpha" switched on.
The first option forces \Space\ to estimate the stellar parameters and chemical
abundances using a loop, whereby the stellar parameters are
derived again using the last abundance estimation, and vice versa, until
they reach convergence.
The second option imposes that the absorption lines of the elements
Mg, Si, Ca, and Ti must be derived as if they were one single element
called the $\alpha$-element \footnote{Other $\alpha$ elements, such as O
and S, are not considered here because these are among the elements
whose abundances derived by \Space\ are (to date) considered not
reliable.} by forcing the relative abundances of these elements to be
equal to each other. Similarly, all of the other non-$\alpha$ elements are
derived as though they were the same element that we call ``Fe". Although
this may appear imprecise, this parameter traces more closely Fe than 
the other elements because most of the absorption lines that drive it 
are in fact iron lines.
The choice of grouping alpha elements and
heavy elements together is due to the features of our spectra. A good fraction
of the LAMOST spectra have low signal-to-noise and their low spectral
resolution does not allow individual fit for most of the
absorption lines. By grouping together absorption lines for similarly
behaving elements, the total absorbed flux is bigger than for the
individual elements and so it makes detectability easier in low
signal-to-noise spectra. The ``lack of purity" for the $\alpha$- and
Fe-abundance is the price we pay to obtain abundance measurements for a
higher number of spectra and with greater precision.

\subsection{Complementary parameters}\label{sec:complementary}
The first few columns of the \Space\ output
report some parameters that, although usually not used for science,
can be helpful in determining features and quality of the
spectra. These parameters are as follows:

\begin{itemize}

\item{\bf FWHM}:
The standard \Space\ pipeline estimates only one FWHM for the
whole spectrum. Because LAMOST spectra are actually composed of two
separate spectra (the blue and the red part), which have different
spectral resolutions, the LAMOST version of \Space\ estimates FWHMs
for the blue and the red spectrum parts independently. These are
output as FWHMb and FWHMr, respectively.
The FWHMb and FWHMr differ and their values are (on average) 
$\sim$2.9 and $\sim$4.3{\AA}, respectively
(see top panel of Fig.~\ref{chi2_FWHM_RV_distr}). Only $\sim$2\% of
the spectra have FWHMb$>$4\AA\ and $\sim$3\% have FWHMr$>$6\AA.
These values may be used as quality selection criteria, although we do
not adopt any such cuts in this study as the number of stars with
very high FWHM is negligible.
Spectra with unusually large FWHM may be out-of-focus spectra, fast
rotator stars, or have bad seeing, for which we do not expect reliable
parameter estimations. As expected, the two FWHMs are weakly
correlated, meaning that if the FWHM is large in the blue arm then
there is a tendency for it to be large for the red arm as well.

\item{\bf Radial Velocity}:
Like the FWHM, the radial velocity correction is also independently 
estimated for the blue and the red parts of the spectrum (RVb and RVr,
respectively). Note that \Space\ was designed to take radial velocity
corrected spectra and, in the case of small errors in the wavelength
calibration or RV correction, it can correct for this by shifting the
central wavelengths of the lines in the model spectrum. The limit of this shift is 1.27FWHM,
which is equivalent to a 3$\sigma$ shift for a Gaussian profile;
this limit corresponds to an RV offset of $\sim\pm$200~\kmsec\ for
an average FWHMb=2.9\AA. Because the FWHM varies for different spectra, the
RV correction limit varies as well.
\footnote{While this work was in progress, we discovered a bug in the
code which can cause the limit of the RV correction to change for different
spectra. This is discussed in Appendix \ref{appendix:bugs}. 
As a consequence of this bug some spectra will fail to converge,
but the parameters reported in the catalog for the converged stars
are unaffected.} For stars having a RV that causes a wavelength shift larger
than 1.27 FWHM, \Space\ quits the analysis reports no results.

Since the LAMOST spectra are corrected for the Earth's motion
only (i.e., they are in the heliocentric frame), the RV correction
performed by \Space\ corresponds to the heliocentric RV of the star. 
This means that the RVb and RVr should agree 
to within their uncertainties. If the difference
between RVb and RVr is large it may indicate a mis-calibration of the
wavelength or an incorrect RV convergence of \Space\ in one of the two
parts of the spectrum. In the bottom panel of Fig.~\ref{chi2_FWHM_RV_distr}
the grey solid histogram shows the distribution of RVb minus RVr for
the first full year of the survey.\footnote{We found that there is a
problem with LAMOST's radial velocity calibration for the red part of the
spectra prior to MJD = 55945 and so here we only plot data from the
first full year of the survey, i.e., from 29th Sept 2012. Note that
the stellar parameters should be unaffected for these early spectra.}
This distribution is not symmetric and peaks at $\sim-7$~\kmsec, which
means that typical spectra have a value of RVr that is 7~\kmsec\ larger than RVb.
This offset remains even if we select only high S/N spectra. The cause of this shift
may be related to the 5.7\kmsec\ difference between LAMOST and APOGEE radial
velocities reported by Tian et al. \citealp{tian} (see section 2.3).
If we compare our radial velocities to those from the LAMOST
pipeline, we find that less than 0.5\% of stars with S/N $>$ 40 have
offsets greater than $\pm$30~\kmsec. From this we conclude that our stellar
parameters are unlikely to be affected by problems with the radial
velocity correction.

\begin{figure}
\centering
\includegraphics[bb= 240 425 390 576]{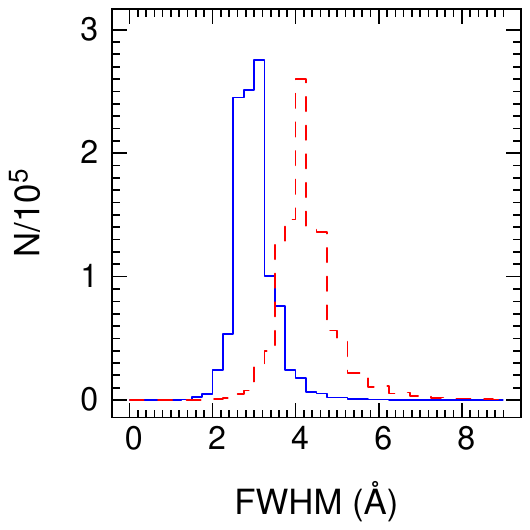}
\includegraphics[bb= 88 425 237 576]{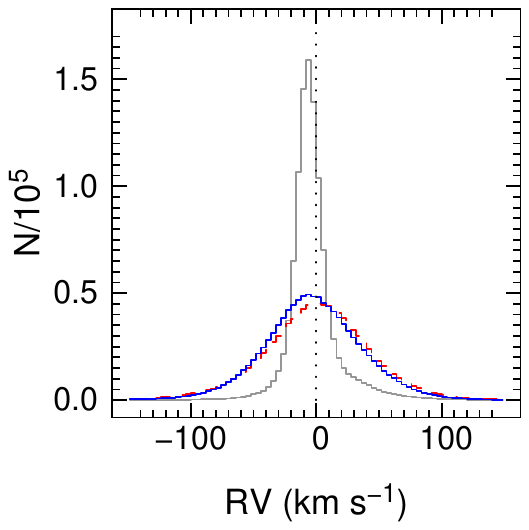}
\caption{Top panel: 
The distribution of the FWHMs for the blue part (blue solid line) and the
red part (red dashed line) of the spectra. Bottom panel: The distribution 
of radial velocities for the blue (RVb) and the
red (RVr) part of the spectra (blue solid and red dashed histograms, respectively). The
solid gray histogram represents the distribution of the difference RVb minus
RVr. For this panel we have only plotted data from the first full year
of the survey (see Section~\ref{sec:complementary} for
details).}
\label{chi2_FWHM_RV_distr}
\end{figure}

\item{\bf S/N}: This is a quality indicator of the spectra.
As discussed above, each spectrum will have a S/N that varies with
wavelength.
\Space\ computes a pixel-by-pixel S/N
based on the discrepancy between the observed spectrum and the best
matching model, using a 50-pixel interval (25-pixel half-width)
centered on the pixel under consideration.
However, in its output \Space\ delivers a single value for the
S/N. This is computed as \SNspace$=1/\sigma_{\rm tot}$, where
$\sigma_{\rm tot}$ is the standard deviation of the residuals between
the observed spectrum and the best matching model over the whole
spectrum, i.e., this can be considered as an average
S/N\footnote{Throughout this paper
we use the term S/N to refer to the signal-to-noise as general meaning and the
term \SNspace\ to refer to the signal-to-noise derived by \Space. When
we refer to the LAMOST signal-to-noise we use the quantity SNRI, which
is calculated in the I-band.}.
In Fig.~\ref{SN_distr} we compare \SNspace\ to the $I$-band
signal-to-noise (SNRI) given in the LAMOST catalog, where we only include spectra that
have stellar parameters in both catalogs. The two S/Ns show a fair 1:1
trend,
although for some spectra the difference can be significant (note that
the SNRI is evaluated in a wavelength window centered on 6220\AA). 
The peak of the S/N distribution
is around 30, with the \Space\ distribution being slightly
higher. This shift is probably due to the aforementioned issue
regarding LAMOST's flux error estimate (see Section~\ref{sec:SPAce-LAMOST}).
For lower S/N spectra one or both methods often fail to converge,
mainly due to a lack of information in the spectra. Other issues that
affect convergence are cosmetic defects
(such as cosmic rays, fringing or dead pixels),
emission lines, or any other
peculiar/unexpected features, like non-stellar objects.
As mentioned in the previous bullet point, \Space\ may also fail to
converge if the radial velocity correction is too large. We return to
the issue of performance later in Section~\ref{sec:comparison_lasp}
when we compare the results of \Space\ to those from the official 
LAMOST pipeline LASP.

\begin{figure}
\centering
\includegraphics[bb=85 423 241 575]{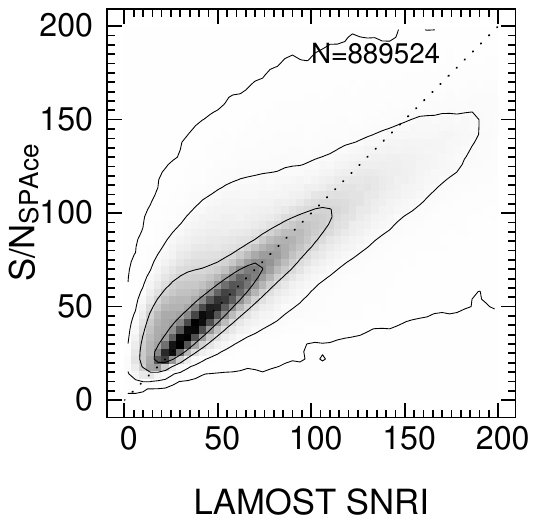}
\includegraphics[bb=240 423 396 575]{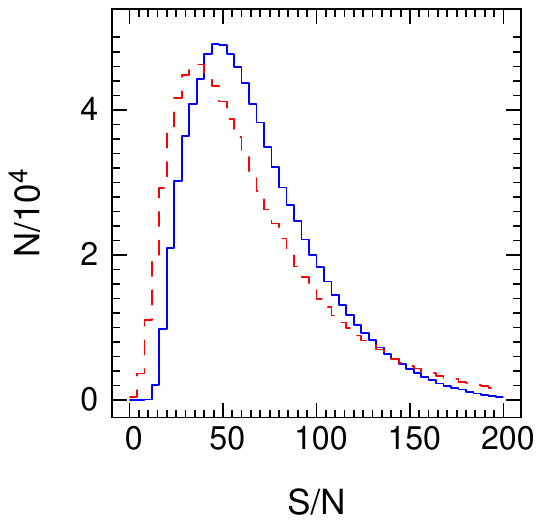}
\caption{Top panel: 1:1 comparison between \SNspace\ and the
LAMOST SNRI (S/N in the I band). Bottom panel: distributions of \SNspace\
(blue solid line) and LAMOST SNRI (red dashed line) for the spectra that have results in both
catalogs. The contours enclose 34, 68, 95, and 99\% of the sample.}
\label{SN_distr}
\end{figure}

\end{itemize}

\subsection{Objects class discrimination}\label{sec:class}
Although the majority of the LAMOST spectra processed with \Space\ are stellar
objects, a fraction of the sample are non-stellar
such as galaxies, quasars, planetary nebulae. Because \Space\ has
been designed to derive stellar parameters of FGK stars only, it
is expected not to converge (and to exit without results) for other objects.
According to the LAMOST 1D pipeline's classification ({\tt CLASS}),
95\% of our sample are stars.
\Space\ converges for 56\% of these, which is a similar success rate
as for the LAMOST pipeline (58\%), and most have good S/N (96\% have
\SNspace$>$20). 
On the other hand, from the 5\% of spectra classified as non-stellar
objects, \Space\ converges for about 1\% of them (1331 objects). A
by-eye inspection shows that these appear to be mostly bona fide
stars, indicating that the LAMOST {\tt CLASS} classification is
not perfect. Note that we are unable to determine the false-positive rate for
{\tt CLASS}, i.e., the fraction of non-stellar spectra that were
classified as stars.

\section{Validation}\label{sec:validation}
To verify the reliability of the LAMOST stellar parameters and chemical
abundances obtained with \Space\, we compared them against
stars with stellar parameters in the literature.
For these tests we restricted ourselves to LAMOST spectra with
sufficient S/N to allow for a fair estimation of the chemical
abundance, so we only selected stars with \SNspace$>$40. We return to the
issue of performance as a function of \SNspace\ later in Section~\ref{sec:uncertainties}.

\begin{figure*}[t]
\centering
\includegraphics[bb=90 486 465 670,width=12cm]{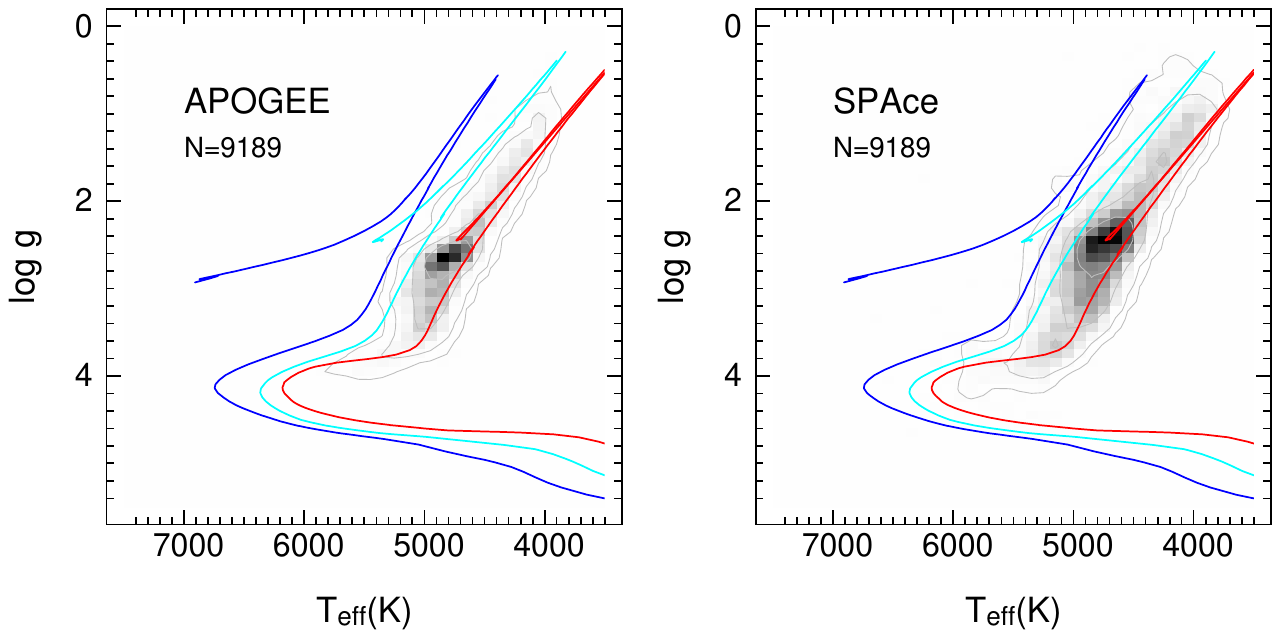}
\includegraphics[bb=35 433 543 570,width=18cm]{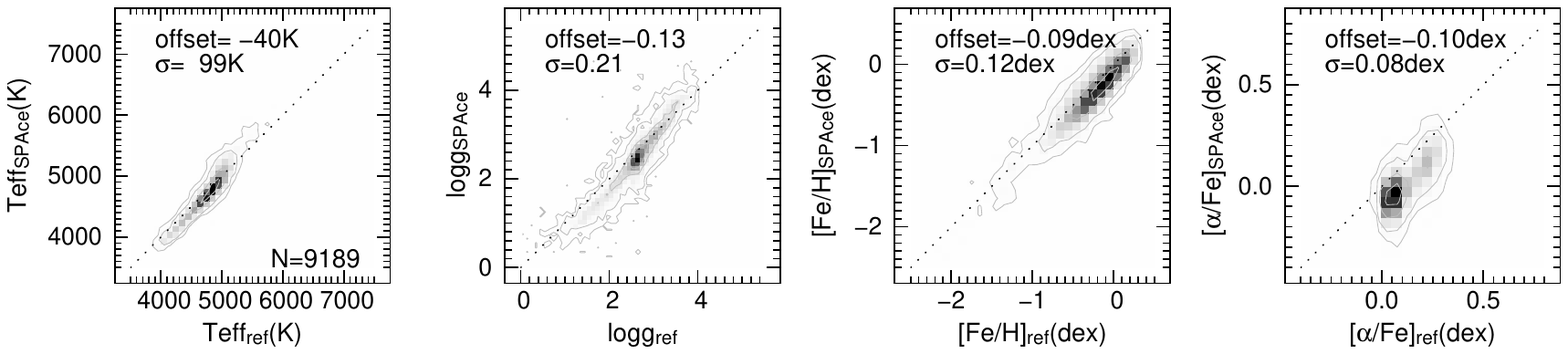}
\caption{An analysis of APOGEE stars with spectra also observed by
  LAMOST. {\bf Top}: Distributions of the stars with ASPCAP calibrated stellar
parameters from the APOGEE spectra (top left panel) and \Space\ stellar parameters
(top right panel) obtained from the LAMOST spectra with \SNspace$>$40. The red, light blue, and
blue lines represent isochrones for [M/H]=0.0~dex (of 5Gyr age), -1.0~dex and -2.0dex
(10Gyr age), respectively. {\bf Bottom}: comparison of
the same \Space\ stellar parameters from the LAMOST spectra to the
reference ASPCAP parameters from the APOGEE spectra. The reference
iron abundance is [Fe/H] (not [M/H]) from ASPCAP. The contours enclose 34, 
68, 95, and 99\% of the sample. A complete 
version of this plot is given in Fig.~\ref{corr_apo_SN40} of the appendix.}
\label{TG_apo_space_calib}
\end{figure*}
\subsection{Comparison to APOGEE}
\label{sec:APOGEE}
APOGEE (Holtzman
et al. \citealp{holtzman}) is a large spectroscopic survey that has
collected over $\sim 150,000$ spectra in the near infrared with a spectral
resolution of $\sim$22\,500. The APOGEE stellar parameters
and chemical abundances are derived with the APOGEE
Stellar Parameters and Chemical Abundances Pipeline (ASPCAP;
Garc{\'i}a P{\'e}rez et al. \citealp{garcia}), which compares the
observed spectra to libraries of theoretical spectra and then
calibrates the resulting parameters and abundances using an observed
calibration sample.
Although method and data are different from LAMOST's, the comparison
is useful.
The complete DR12 APOGEE sample has 34\,783 stars in common with
LAMOST (45\,193 LAMOST spectra, including repeat observations). 
Out of these spectra, we ignored the 27\,021 spectra which are flagged
by APOGEE as having possible problems in the spectrum (for instance, the
stars may have a bright neighbor that can pollute the spectrum,
particularly broad lines or low S/N that can badly affect the
parameter derivation)\footnote{In our APOGEE sample
all the spectra having [Fe/H]$<-0.6$~dex were flagged as problematic by
Holtzman et al. during their abundance estimation. 
Since we wish to retain the comparison with these stars 
we neglected such flags.}.
This left us with 18\,172 stellar spectra that have no reported
problems, of which 13\,351 are giant stars (\logg$<$3.5) with calibrated stellar
parameters. We neglect the dwarf stars because APOGEE does not provide
reliable parameters for these stars (Holtzman et al., \citealp{holtzman}).
\Space\ converged for 10\,879 spectra of those giant stars, of which 9\,189 had
\SNspace\ above our aforementioned threshold of 40.

The performance of the two catalogs is shown in
Fig.~\ref{TG_apo_space_calib}. The top panels show the 
(\temp,\logg) plane, together with a set of fiducial isochrones
by Bressan et al. \cite{bressan} corresponding to typical thin-disc,
thick-disc and halo populations.
\Space\ provides a good match to the expected distribution, as
demonstrated by the agreement with the isochrones.
In the bottom panels of Fig.~\ref{TG_apo_space_calib} we directly
compare the \Space\ and APOGEE stellar parameters, while a more detailed
plot showing the correlations between the parameters is given in the
Appendix \ref{appendix:plots} (Fig.~\ref{corr_apo_SN40}).
Since APOGEE gravities are calibrated using asteroseismic
gravities from Kepler, we defer the discussion of the gravity
performance to the following section where we directly compare to data
from Kepler (Section~\ref{sec:kepler}).
The distributions show some systematic differences (shifts and/or
correlations) between APOGEE and \Space\ parameters. 
When making this comparison we should first point out that prior tests
of the \Space\ pipeline have shown a tendency to slightly
underestimate [Fe/H], at a level of around $-0.05$~dex (see Fig.~17 
and accompanying discussion in
Boeche \& Grebel \citealp{boeche}). We find that \Space\
[Fe/H] and [$\alpha$/Fe] abundances appear to be underestimated with
respect to APOGEE's. However, for both parameters the offset appears
to be independent of [Fe/H] or [$\alpha$/Fe], which means that it is
straight-forward to shift the \Space\ abundances onto the APOGEE scale. 
A closer look (Fig.~\ref{corr_apo_SN40}) shows that the 
underestimation in [$\alpha$/Fe] has a very weak correlation with
the other parameters.
[Fe/H], on the other hand, does show stronger correlations
(most notably with gravity and temperature), but the offset appears to be
independent of [Fe/H] itself.

\begin{figure*}
\centering
\includegraphics{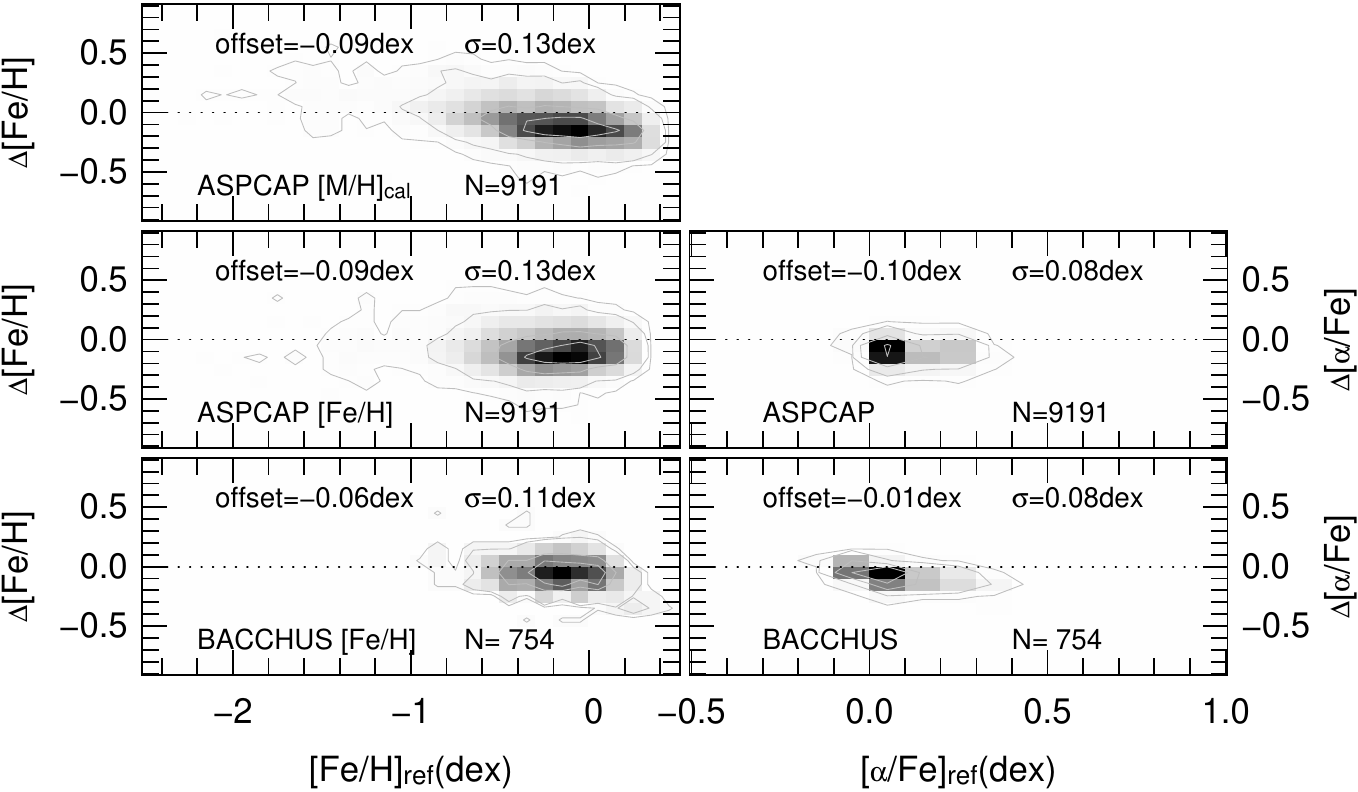}
\caption{Residuals between \Space\ and a selection of
different reference abundances available for the APOGEE data
(\Space\ minus reference value) for stars with \SNspace$>$40. The top
panel shows the ASPCAP [M/H], the middle panel the ASPCAP [Fe/H] and
[$\alpha$/Fe], all of which are described in Holtzman et
al. \citealp{holtzman}). The lower panel shows [Fe/H] and
[$\alpha$/Fe] from the BACCHUS pipeline (Hawkins et
al. \citealp{hawkins}). The horizontal axes denote the reference
abundance for that particular panel. The contours enclose 34, 
68, 95, and 99\% of the sample.}
\label{fig:bacchus}
\end{figure*}

Note that in the above discussion we have used the ASPCAP iron
abundance [Fe/H], not the total metallicity [M/H] (PARAM\_M\_H). The
latter quantity has been calibrated onto an [Fe/H]-scale using
literature abundances for star clusters, but the former has not (see
section~5.4.3 of Holtzman et al. \citealp{holtzman}). The lack of an
external calibration may influence our findings and so we check the
comparison between \Space\ and ASPCAP [M/H] in
Fig. \ref{fig:bacchus}. For completeness we also include [Fe/H] and
[$\alpha$/Fe] from the BACCHUS pipeline (Hawkins et
al. \citealp{hawkins}), which provides a complementary abundance
analysis using APOGEE stars with asteroseismic gravities from Kepler. For
the BACCHUS [$\alpha$/Fe] we take the average of the Mg, Si, Ca and Ti
abundances. We can see that when using the ASPCAP [M/H] there exists a
noticeable correlation, which may be due to uncertainties in the
external calibration of [M/H] at the metal-poor end. The correlation
disappears (but an offset remains) when comparing \Space\ to both
ASPCAP [Fe/H] and BACCHUS [Fe/H]. The only differences between BACCHUS
and ASPCAP [Fe/H] is a marginal reduction in both the scatter and
offset in the former. For [$\alpha$/Fe] the results are 
also similar between BACCHUS
and ASPCAP, although \Space\ appears to exhibit a very slight
correlation in [$\alpha$/Fe] compared to BACCHUS.

We include three further plots in Appendix
\ref{appendix:plots}. Fig. ~\ref{plot_apo_bacchus_met_bin} shows
the (\temp,\logg) distributions for stars measured by ASPCAP,
BACCHUS, and \Space, divided into bins of [Fe/H]. For each bin we have
over plotted isochrones for the corresponding metallicity. The
placement of the \Space\ red clump is slightly more consistent with
the corresponding isochrones, compared to both APOGEE/ASPCAP and
BACCHUS. Furthermore, at lower metallicities ([Fe/H] $\lesssim-0.4$) the
\Space\ red giant branch appears to match the isochrones better than
ASPCAP.\\
The [$\alpha$/Fe] performance is characterized in
Figs. \ref{apo_space_aFe}  and \ref{comp_alphaFe_apo} of Appendix
\ref{appendix:plots}. The former shows the distribution of stars in
the ([Fe/H],[$\alpha$/Fe]) plane for \Space\ and ASPCAP, while the
latter shows a one-to-one comparison for bins of [Fe/H].
The low- and high-alpha sequences are correctly estimated by \Space,
in that \Space\ detects a clear shift in alpha between stars from
the low- and high-alpha sequences (Fig.~\ref{comp_alphaFe_apo}).
Although the LAMOST alphas are not as precise as those from APOGEE
(e.g. there is no visible gap between the two sequences),
the performance of \Space\ is impressive considering the fact that the
LAMOST data have a 10 times lower resolution.

\begin{figure*}[t]
\centering
\includegraphics[bb=90 486 465 670,width=12cm]{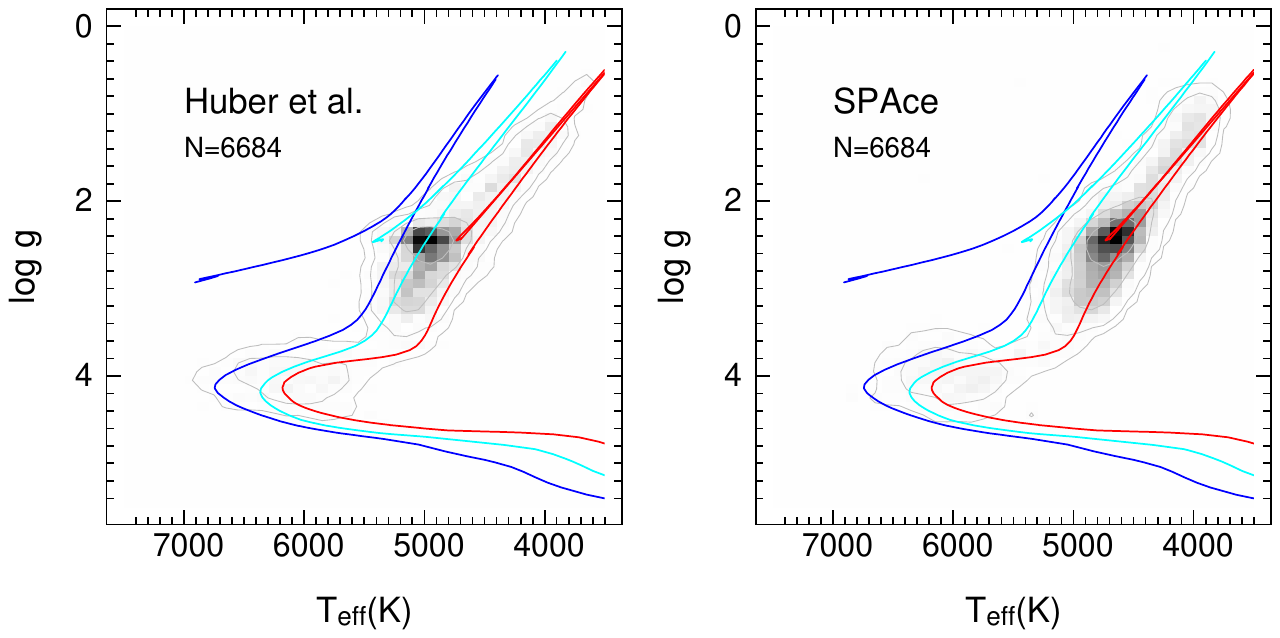}
\includegraphics[bb=35 433 422 562,width=16cm]{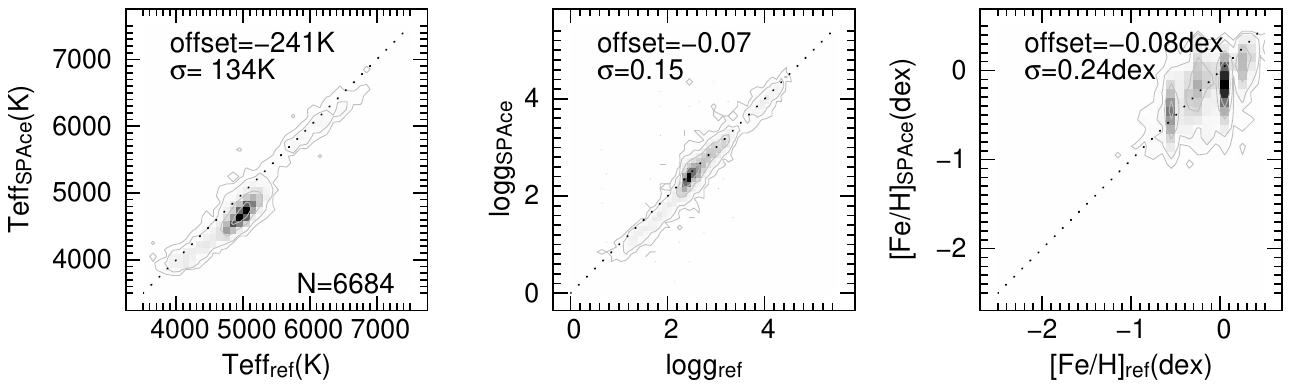}
\caption{{\bf Top}: distributions in the (\temp, \logg) plane of the
Kepler stars as given by Huber et al. (left panel) and \Space
(right panel), where we have restricted ourselves to stars with \SNspace$>$40. 
The red, light blue, and blue lines represent the same
isochrones as in Fig.~\ref{TG_apo_space_calib}. {\bf Bottom}:
comparison of the \Space\  stellar parameters to the reference Huber
et al. parameters.  The contours enclose 34, 68, 95, and 99\% of the
sample. A complete version of this plot is given in
Fig.~\ref{corr_huber_SN40} of the appendix.}
\label{TG_huber_space}
\end{figure*}

\subsection{Comparison with Kepler stars}\label{sec:kepler}
Huber et al. \cite{huber} report the properties of $\sim$20\,000 stars
observed by the the NASA Kepler mission. While the stellar parameters of
most of these stars come from a collection of different catalogs with different
observational techniques (photometry, spectroscopy), $\sim$15\,500 of these
stars have known oscillations that permit a precise measurement of their
surface gravity through asteroseismology. These stars can therefore be
used as reference in \logg\ for comparison purposes.
Our LAMOST sample has 6684 stars in common with Huber et
al. \cite{huber}. Most of these are giant stars, with a small number
of sub-giant and turn-off stars. In Fig.~\ref{TG_huber_space} we show the distribution of
these stars in the (\temp, \logg) plane along with a direct comparison of the
parameters, while in Fig.~\ref{corr_huber_SN40} of the Appendix we show the
correlations between stellar parameters and the residuals with respect to the
reference. In these figures we see an apparent offset in \temp\ and a poor match in
[Fe/H] between \Space\ and the reference values. The temperature offset
appears to be in agreement with the one found by Huber et al. 
(\citealp{huber}; top panel of Fig. 7), namely a systematic 
overestimation in their \temp\ value due to the limitations of the Kepler
Input Catalogue (KIC) from which they were obtained.
Similarly the metallicity offsets are not important, since the Huber
et al. \cite{huber} metallicities, which are mainly derived from the
KIC, are known to have been underestimated (e.g., Dong et al. \citealp{dong}).
Furthermore the KIC metallicities show an unnatural discrete
distribution, which demonstrates their lack of precision.
We do know that Huber's \logg\ values are extremely accurate (typically
0.03 dex), and the central bottom panel of Fig.~\ref{TG_huber_space}
(or the middle row of panels in Fig.~\ref{corr_huber_SN40} of the
Appendix) shows a very good  match with the \Space\ gravity, albeit
with an overestimation of $\sim$0.2 when \logg$\lesssim$2 dex.

\begin{figure}
\centering
\includegraphics[bb=94 310 465 670,width=\hsize]{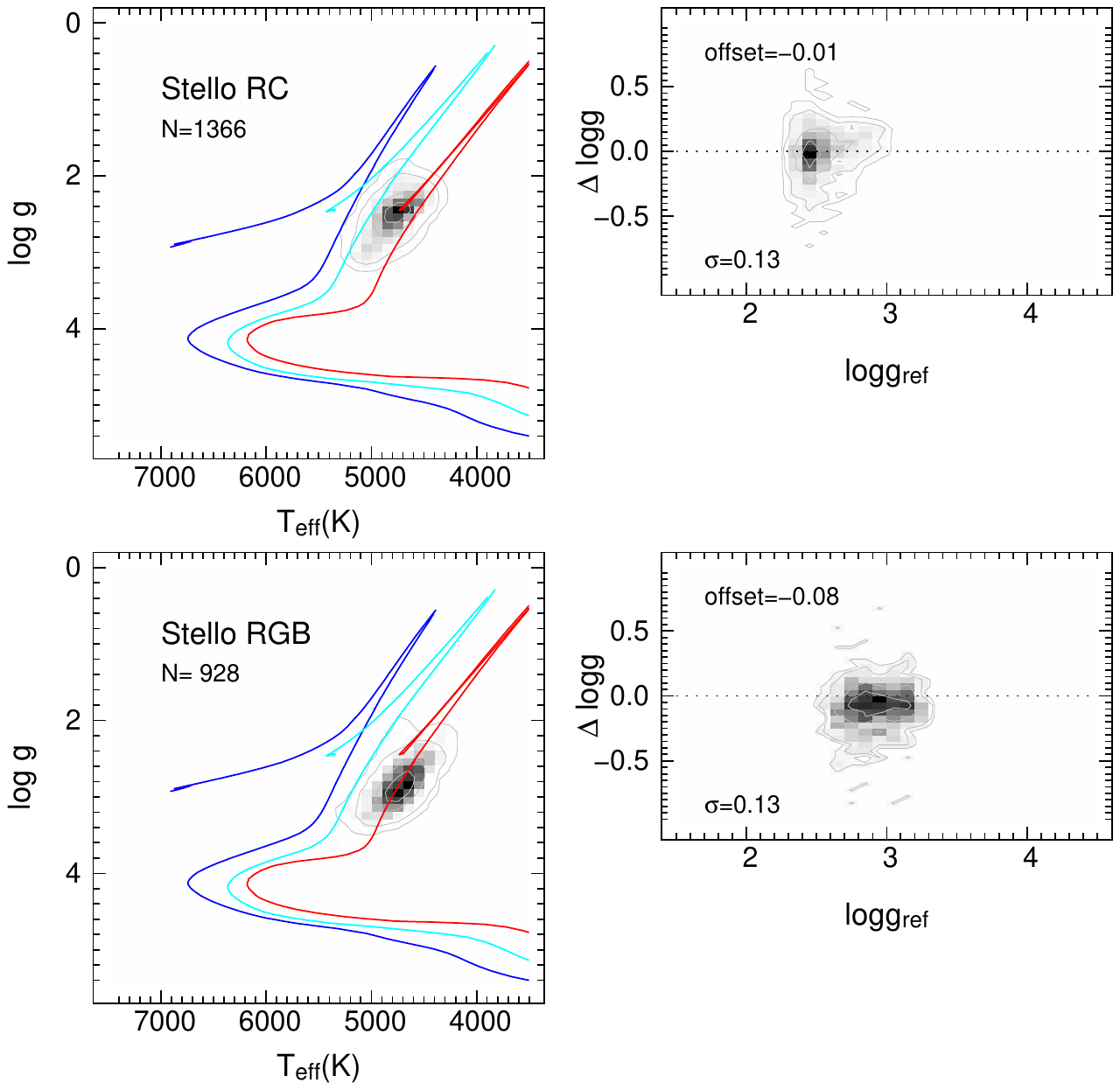}
\caption{
Top panels: comparison of \Space\ to seismic
gravities from the updated Stello et al. (\citealp{stello}) sample
(see Section~\ref{sec:kepler}) for 1366 red clump
stars for stars with \SNspace$>$40. The left panel shows the
distribution of \Space\ parameters in the  
(\temp, \logg) plane, while the right panel shows the \logg\ residuals
between \Space\ and Stello.
Bottom panels: As above, but for 928 red giant stars.}
\label{fig:stello}
\end{figure}

Stello et al. \citep{stello} carried out a further analysis of this Kepler
sample to distinguish red clump (RC) and red giant branch (RGB) stars
following a method similar to Bedding et al. \cite{bedding}. The
analysis by Stello et al. \cite{stello} included only two years
of Kepler data. To refine those results we repeated their analysis on
3.5 years of Kepler data.  The comparison with these updated seismic
results is presented in Fig~\ref{fig:stello}. In the left panels we
notice that \Space\ correctly positions both the RC and RGB locus on
the high metallicity isochrone. The right panels show that \Space\ is
able to recover accurate gravities for both the RC and RGB samples,
although the latter have a slight systematic offset ($\sim -0.08$ dex).
It should be stressed that these results are obtained directly from
\Space\ without any need for calibrations.

\subsection{Main-sequence stars: Comparison with the Geneva-Copenhagen
  and Gaia-ESO surveys}
\label{sec:dwarfs}

As the previous sections focused mainly on giant stars, we now
utilize two surveys that contain large numbers of main-sequence
stars.
Our first data set is from the Geneva-Copenhagen Survey (GCS,
Nordstr\"om et al. \citealp{nordstrom04}), which consists of 16\,682
FGK dwarf stars in the immediate solar neighborhood. As reference
parameters we adopt those of Casagrande et al. \citep{casagrande}. In
this work they derived \temp\ and \logg\ from photometry using an
infrared flux method (IRFM), where the latter quantity also
incorporated the Hipparcos parallax, and derived metallicity from
Str\"omgren photometry. Although we cannot expect high precision from
this approach, their parameters are still of great help in determining the
robustness of our parameters for dwarf stars. We have 413 stars in
common with GCS, out of which 233 have \SNspace$>$40 and \Space\ stellar
parameters. Their distributions in the (\temp, \logg) plane and a
direct comparison are shown in Fig.~\ref{TG_GCS_space}, while the full
correlations between the parameters are in Fig.~\ref{corr_GCS_SN40} of
the Appendix.

Our second data set is from the Gaia-ESO survey (GES, Gilmore et
al. \citealp{gilmore}), which provides precise stellar parameters from
high resolution spectra obtained with the ESO UVES and Giraffe
spectrographs (with a spectral resolution of $\sim$60\,000 and 
$\sim$20\,000, respectively).
Using GES DR3 we have found 390 stars in common with
LAMOST that have non-null GES parameters \temp, \logg, and [Fe/H]. Out
of these, 146 have \SNspace$>$40 and \Space\ stellar
parameters.
Comparisons between \Space\ and GES are presented in
Fig.~\ref{TG_GES_space} and in Fig.~\ref{corr_GES_SN40} of the Appendix.

For both the GCS and GES comparisons we can see that \temp\ and
\logg\ match the reference parameters fairly well, with small
offsets. However, for the GCS the iron abundance exhibits a negative offset
(i.e. \Space\ is underestimating [Fe/H] compared to GCS), especially for the hotter,
lower-gravity dwarfs (see upper-left and upper-middle panels of
Fig.~\ref{corr_GCS_SN40}). A handful of these stars also have
significantly under-estimated gravities compared to the GCS
value. There is weak evidence for a similar metallicity trend in the
GES comparison (Fig.~\ref{corr_GES_SN40}), although 
the paucity of stars in this region of parameter space
(i.e., \temp\ $\gtrsim$ 6000~K and 3 $\lesssim$ \logg\ $\lesssim$ 4
dex) makes it hard to draw any firm conclusions.
From Fig.~\ref{corr_GES_SN40} we can see that there is an
underestimation in [$\alpha$/Fe] of $\sim$0.1~dex when
comparing \Space\ to GES, which is similar to the one seen when 
comparing \Space\ to APOGEE (Section~\ref{sec:APOGEE}).

\begin{figure*}[t]
\centering
\includegraphics[bb=90 486 535 680,width=12cm]{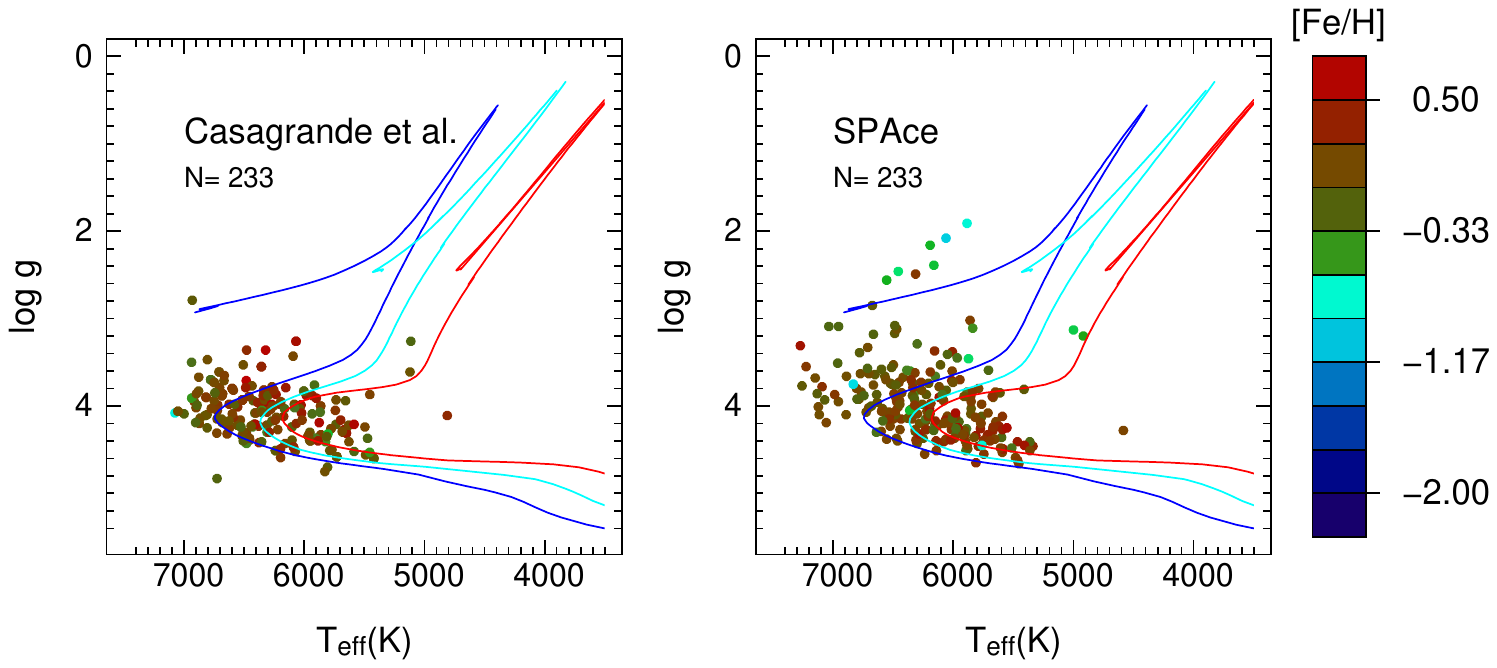}
\includegraphics[bb=36 433 422 565,width=16cm]{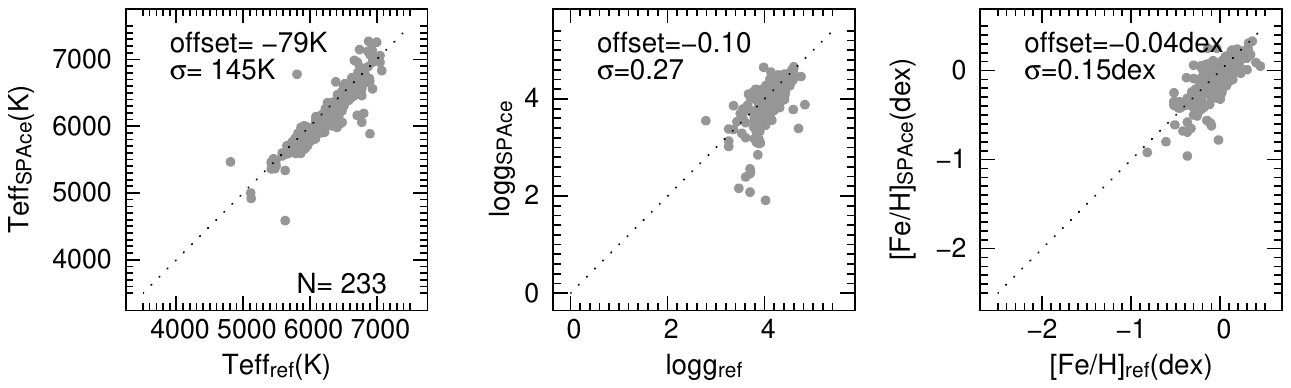}
\caption{{\bf Top}: distributions in the (\temp, \logg) plane of the
GCS stars as given by Casagrande et al. stars (left panel)
and \Space\ (right panel) for stars with \SNspace$>$40. The
red, light blue, and blue lines represent the same isochrones as
in Fig.~\ref{TG_apo_space_calib}. {\bf Bottom}: comparison of
the \Space\ stellar parameters to the reference Casagrande et
al. parameters. A complete version of this plot is given in
Fig.~\ref{corr_GCS_SN40} of the appendix.} 
\label{TG_GCS_space}
\end{figure*}

\begin{figure*}[t]
\centering
\includegraphics[bb=90 486 535 680,width=12cm]{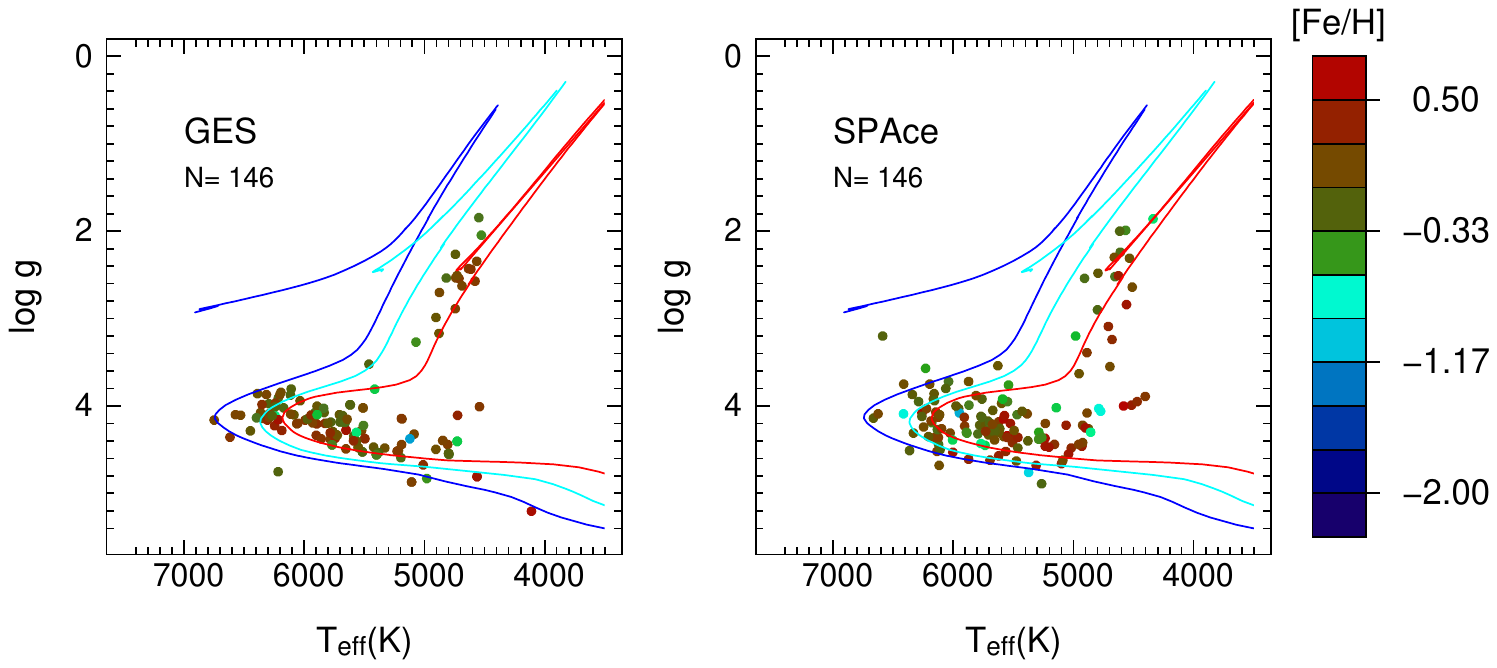}
\includegraphics[bb=35 433 543 570,width=16cm]{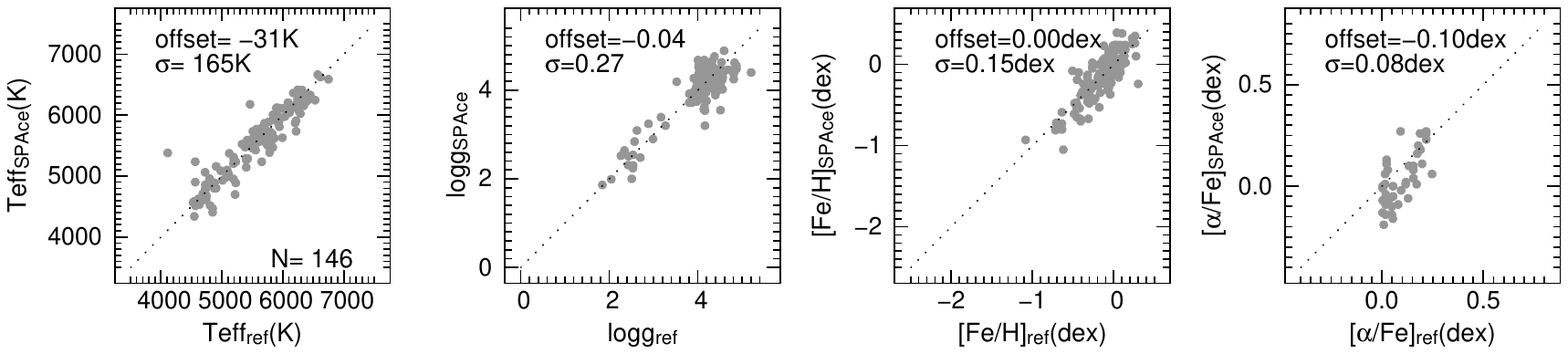}
\caption{{\bf Top}: distributions in the (\temp, \logg) plane of the stars as given by
the GES stars (left panel) and \Space\ (right panel) for stars
with \SNspace$>$40. The red, light blue, and
blue lines represent the same isochrones as in Fig.~\ref{TG_apo_space_calib}. {\bf Bottom}: comparison
of the \Space\ stellar parameters to the reference GES parameters.
A complete version of this plot is given in Fig.~\ref{corr_GES_SN40} of the appendix.}
\label{TG_GES_space}
\end{figure*}


\subsection{Uncertainties}\label{sec:uncertainties}

\Space\ estimates $1\sigma$ uncertainties along with the stellar
parameters for each star. It does this by considering the shape
of the $\chi^2$ hyper-surface and identifying the upper and lower
parameter limits for which $\chi^2$=$\chi^2_{min}+\Delta\chi^2$, where
$\Delta\chi^2$ depends on the number of degrees of freedom (for
details see section 7.6 in Boeche \& Grebel, \citealp{boeche}).
For a correct validation of these uncertainties 
we would need to compare the \Space\ results to reference parameters having no 
(or very small) errors, which would allow us to infer the precision
from the standard deviation of the discrepancies and the accuracy
from the systematic offsets.
Unfortunately this is not possible because the reference parameters
are also affected by stochastic and systematic errors. However, we can
overcome this by adding in quadrature the errors from
both \Space\ and the reference data set, although of course this
relies on the assumption that the reference errors are accurately
reported.

The comparison between the distribution of the discrepancies (\Space\ minus
reference parameter) and the \Space\ errors 
for giant stars is reported in the three top left panels of 
of Fig.~\ref{discr_SN_errors_giants}.
As expected, the discrepancies become larger as the \SNspace\ decreases. The
red dots show the expected magnitude of the standard deviation, as
estimated by adding in quadrature both the \Space\ and ASPCAP
uncertainties.
For most of these parameters the \Space\ errors look like they are
providing a fair approximation to the observed scatter, with the red dots
agreeing with the 1$\sigma$ limit of the distribution of discrepancies
(indicated with black error bars). 
For \temp\ and
[$\alpha$/Fe] \Space\ appears to be overestimating the uncertainties
slightly, especially for [$\alpha$/Fe] in the low-S/N regime. On the
other hand, it appears to underestimate the errors for
[Fe/H]. The magnitude of this underestimation is hard to quantify
because the reported errors on the ASPCAP [Fe/H] are exceedingly small
(0.03 dex) and so it is unclear which method is underestimating the
size of their errors. As an example, if we consider moderately low \SNspace\
stars (around 40 $\lesssim$ \SNspace\ $\lesssim$ 70), we find that in order to bring
our estimated errors into agreement with the standard deviation of the
discrepancy, we would need to inflate both \Space\ and ASPCAP errors
by around 30\%. Repeating this exercise for [$\alpha$/Fe] indicates
that for the same stars we would need to reduce the \Space\ errors by
around 40\% in order to match the observed standard deviation.

In the top right panel of Fig.~\ref{discr_SN_errors_giants} we compare
the \Space\ \logg\ to the Huber et al. values. Because of the high
precision of the seismic gravities (with typical uncertainties of 0.03 dex), we expect
that the scatter reflects the magnitude of the \Space\ errors.  Apart
from a slight overestimation at the low-\SNspace\ end, the \logg\ errors
provide a good match to the observed scatter. However, if we look at
the mean of the discrepancy (given by the black dots) there appears to
be a slight correlation with \SNspace; as well as the previously identified offset
(see Section~\ref{sec:kepler}), the \Space\ \logg\ appears to suffer
from a larger underestimation for lower \SNspace\ spectra. At higher
\SNspace\ ($\sim$100) the
offset is barely noticeable, amounting to less than 0.05 dex, but this increases
to as much as 0.15 dex at lower \SNspace\ ($\sim$30).
The same analysis is presented on the bottom panels of
Fig.~\ref{discr_SN_errors_giants} for the
Gaia-ESO survey stars. This sample is more representative of dwarf stars,
since it contains only a small fraction of giants.
Although there are also far fewer stars to compare to, and hence the
distributions (especially \temp) are a little harder to interpret,
the trends we saw for giants are replicated. Namely our errors appear
to be reliable for \temp\ and \logg, but are slightly over- and
under-estimated for [Fe/H] and [$\alpha$/Fe], respectively. The
correlation between \Space\ \logg\ and \SNspace\ that was found for the Huber
sample appears to persist for the GES, implying that this is common for
dwarf stars as well.


\section{The catalog}\label{sec:catalog}
Out of the 2\,052\,662 processed spectra, \Space\ derived parameters for 1\,097\,231
spectra. For half of the spectra \Space\ did not output any results since it
failed to converge for one or more parameters. The main reason
for the failure of convergence is the low S/N: for spectra with SNRI$<$20
(736\,929), $\sim$80\% failed to converge, while for spectra with SNRI$>$20
(1\,315\,733), $\sim$27\% failed to converge. As we will show later
(Section~\ref{sec:comparison_lasp}), this problem is not unique
to \Space; in fact, the fraction low-S/N spectra that can be processed by
\Space\ is similar to that of
the official LAMOST pipeline (LASP). Other reasons for failure can be: stars
with stellar parameters out of the limits of the GCOG library
(e.g., too hot or too cool); too high a RV; spectra of non-stellar, or
peculiar objects.

The final results are sorted and described as in
Table~\ref{table:description}. In this table we include the parameter
{\met}, which has not previously been mentioned.
This parameter represents the metallicity of the atmosphere model that was used
to compute the GCOGs, which were in turn used to derive the spectrum's
parameters. These atmosphere models were taken
from ATLAS12 by Castelli and Kurucz, \citealp{castelli}) and have
[$\alpha$/Fe]=0. This means that their nominal metallicity is equal to
the iron abundance. Since iron is the main driver of the
metallicity, one finds that in our catalog the metallicity \met\ is 
usually very close (or equal) to [Fe/H]. To obtain a more general
metallicity index that also includes the contribution of the $\alpha$ elements,
we suggest using the following formula\\
\begin{equation}
[M/H]^{chem}=\mbox{[Fe/H]}+ \log(0.638\cdot10^{\mbox{[$\alpha$/Fe]}}+0.362),
\end{equation}
given by Salaris et al. \cite{salaris}.

We remember that our [Fe/H] parameter represents the abundance of all
non-$\alpha$-elements as if they were a single element, to which iron is
the major contributor. The total number of absorption lines measured to
derive the abundance of the [Fe/H] parameter is reported by the {\it Nlin}
parameter.
Consequently, the abundances will be more robust for spectra with
higher values of {\it Nlin}.
The same applies to the \aFe\
parameter, which represents the abundance of the $\alpha$-elements Mg,
Si, Ca, and Ti derived as if they were one single element.\\

\begin{figure*}[t]
\begin{minipage}[h]{18cm}
\centering
\includegraphics[bb=94 483 463 670,width=14cm]{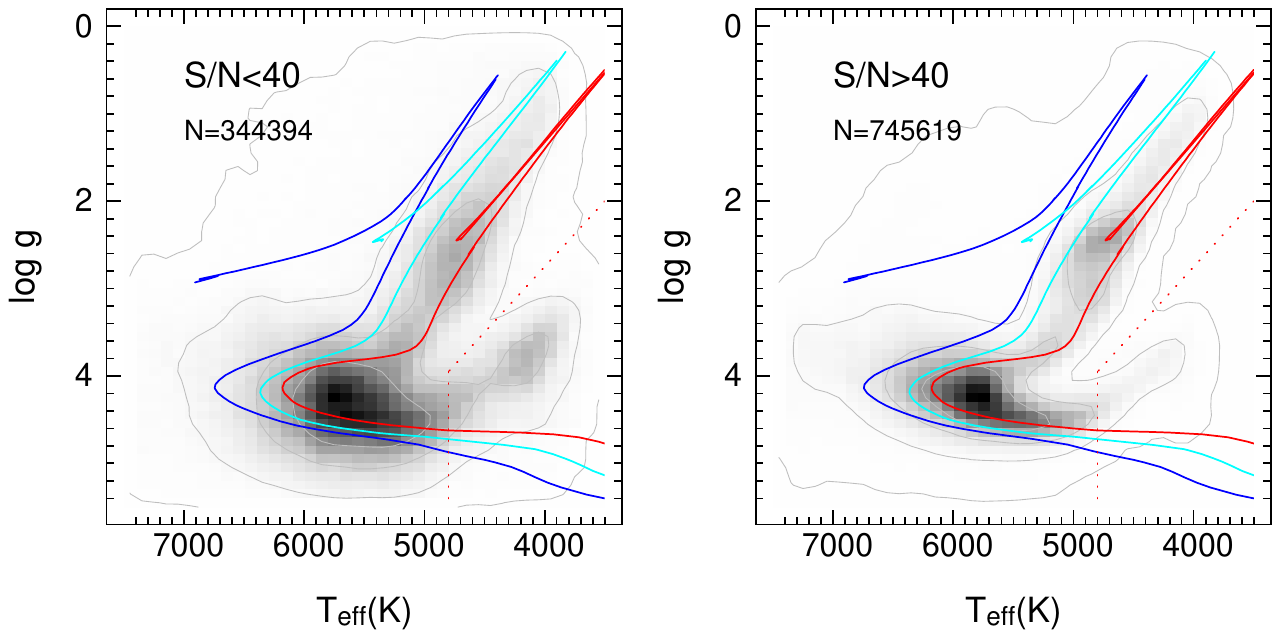}
\caption{Distributions of \Space\ parameters (\temp, \logg) for the LAMOST
DR1 stars with \SNspace\ smaller and larger than 40. The red, light blue, and
blue solid lines represent the same isochrones as in Fig.~\ref{TG_apo_space_calib}. The cool
dwarfs delineated by the red dotted line appear to be subject to a
systematic error in gravity (see Section~\ref{sec:systematics}).
The contours enclose 34, 68, 95, and 99\% of the sample.}
\label{TG_DR1_space}
\end{minipage}
\begin{minipage}[h]{18cm}
\centering
\includegraphics[bb=95 284 480 385,width=14cm]{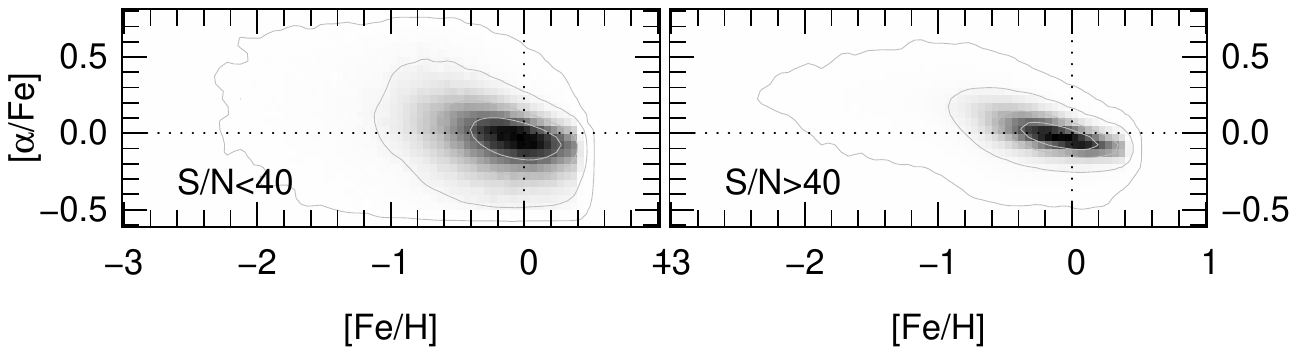}
\caption{Distributions of \Space\ abundances ([Fe/H], [$\alpha$/Fe]) for the LAMOST
DR1 stars with \SNspace\ smaller and larger than 40. 
The contours enclose 34, 68, 95, and 99\% of the sample.}
\label{aFe_DR1_space}
\end{minipage}
\begin{minipage}[h]{18cm}
\centering
\includegraphics[bb=77 575 545 730,width=14cm]{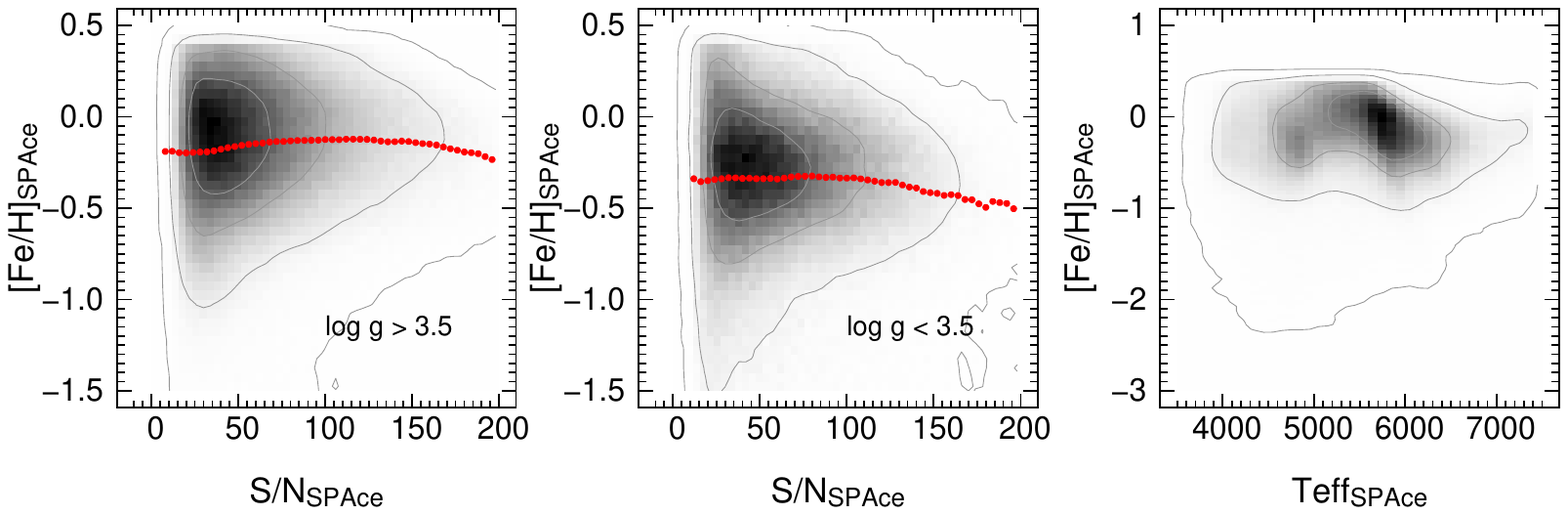}
\caption{Density distributions of [Fe/H] as a function of \SNspace\ for dwarf
(left panel) and giant stars (middle panel). The red dots trace the average
of the [Fe/H] distribution for each S/N bin. Right panel: density
distribution of [Fe/H] as a function of \temp\ for the whole
sample. The contours enclose 34, 68, 95, and 99\% of the sample.}
\label{system_err_space}
\end{minipage}
\end{figure*}

In Figs.~\ref{TG_DR1_space} and \ref{aFe_DR1_space}
we show the distributions of
\logg\ vs \temp\ and [$\alpha$/Fe] vs [Fe/H] for
our \Space\ catalog, divided into high and low \SNspace. 
We here employ \SNspace=40 as a discriminator between the stellar parameters
with ``fair" and ``good" accuracy and, as expected, the dispersion of
the data observed in these figures reflects this.
In Fig.~\ref{TG_DR1_space} the red dotted line delineates a sample of
cool dwarf stars that, as they are placed far from the expected isochrones,
are most likely subject to a systematic error in gravity. This is
discussed in the following section.

\begin{deluxetable*}{lcccl}
\tablecaption{Description of the catalog}
\tablecolumns{5}
\tablenum{1}
\tablewidth{0pt}
\tablehead{
\colhead{Field} &
\colhead{Name} &
\colhead{Format} &\colhead{Unit} & \colhead{Description} \\
}
\startdata
1  & OBSID & integer & ... & unique identification number of the spectrum \\
2  & designation & string & ... & object designation \\
3  & ra & float & deg & object Right Ascension \\
4  & dec & float & deg & object Declination \\
5  & obsdate & ... & ... & date of observation \\
6  & lmjd & integer & ... & local modified Julian date \\
7  & planid & string & ... & plan ID in use \\
8  & spid & integer & ... & spectrograph ID \\
9 & fiberid & integer & ... & fiber ID of object \\
10 & RVb & float & km s$^-1$ & radial velocity correction of the blue part of the spectrum \\
11 & RVr & float & km s$^-1$ & radial velocity correction of the red part of the spectrum \\
12 & FWHMb & float &  \AA\ &Full-Width-Half-Maximum of the blue part of the spectrum \\
13 & FWHMr & float &  \AA\ &Full-Width-Half-Maximum of the red part of the spectrum \\
14 & S/N & float & ... & signal-to-noise as computed by \Space  \\
15 & \temp\ & float & K & effective temperature \\
16 & inf & float & K & effective temperature lower limit\\
17 & sup & float & K & effective temperature upper limit\\
18 & \logg\ & float & dex & gravity \\
19 & inf & float & dex & gravity lower limit\\
20 & sup & float & dex & gravity upper limit\\
21 & \met\ & float & dex & nominal metallicity of the atmosphere model \\
22 & inf & float & dex & nominal metallicity lower limit\\
23 & sup & float & dex & nominal metallicity upper limit\\
24 & [Fe/H] & float & dex & iron abundance \\
25 & inf & float & dex & iron abundance lower limit\\
26 & sup & float & dex & iron abundance upper limit\\
27 & Nlin & integer & ... & number of absorption lines used to derive the iron abundance\\
28 & [$\alpha$/H] & float & dex & $\alpha$-element abundance\\
29 & inf & float & dex & [$\alpha$/H] lower limit\\
30 & sup & float & dex & [$\alpha$/H] upper limit\\
31 & Nlin & integer & ... & number of absorption lines used to derive
[$\alpha$/H] \\
32 & flag & integer & ... & spectra with FLAG=0 must be rejected or
treated with extreme caution (Section~\ref{sec:systematics})\\
33 & class & string & ... & LAMOST classification of the object (Section~\ref{sec:class})\\
34 & z & float & ... & red shift of object (from the LAMOST pipeline)\\
35 & z\_err & float & ... & red shift error (from the LAMOST pipeline) \\
\enddata
\label{table:description}
\end{deluxetable*}

\subsection{Known systematic errors}\label{sec:systematics}

In Section~\ref{sec:validation} we demonstrated the successful performance
of \Space\ and found that there is good agreement with reference
data sets. We also pointed out the presence of some
systematic errors. The comparison with the Kepler stars showed that \Space\
underestimates the gravity ($\sim-0.2$) for giants with \logg$\lesssim$2, 
while the comparison with the GCS
and GES stars revealed an underestimation of the iron abundance for stars with
\temp\ $\gtrsim$ 6000~K and 3 $\lesssim$ \logg\ $\lesssim$ 4 (see also later in
this section).
Still, we did not test the
performance for cool dwarf stars because of a lack of reference stars
in this \temp-\logg\ region.
As was shown in
Fig. \ref{TG_DR1_space}, it appears that \Space\ underestimates
gravity for the cool dwarfs. These have been delineated by the red
dotted line in this figure.
The cause is unclear, but may be due to one or more of the following
reasons: the physics of the adopted 1D atmosphere models is deficient,
i.e., does not match the real conditions; the assumption of LTE is
invalid; and/or the presence of molecular lines (which are neglected
by \Space) are affecting the atmospheres of cool dwarfs.
Although this manifests itself in an underestimation in gravity, the
problem could propagate into the other parameter estimates as
well. Therefore we recommend that 
the parameters for these cool dwarfs are not to be trusted. Since we
have no way to determine precisely where this systematic becomes
significant, the red dotted boundary in Fig.~\ref{TG_DR1_space} was
determined by eye.
These are the stars for which \temp$<4800$~K and
\logg$>0.0015\cdot$\temp-3.25 and they are flagged with {\tt FLAG}=0 in the
catalog. They represent 3\% of the total sample.

Additional potential systematic errors are shown in
Fig.~\ref{system_err_space}. The left and middle panels appear to show
a dependence on the mean [Fe/H] with \SNspace, which has an amplitude of
$\sim0.1$ dex, for both dwarfs (left) and giants (middle).
Although this may be due to some feature of the survey selection
function (e.g., the distribution of latitude or distance), we have
been unable to pinpoint a cause and therefore conclude that this is
likely to be a systematic inherent to \Space.
The right panel of Fig.~\ref{system_err_space} shows our final
potential systematic, which has already been discussed above in
Section~\ref{sec:dwarfs}. In that section we pointed out that hot
(\temp\ $\gtrsim$ 6000~K) dwarfs appear to have underestimated [Fe/H].
The same problem appears to manifest itself in the right panel of
Fig.~\ref{system_err_space}, where there is a clear deficiency of hot,
iron-abundant stars, i.e., for these stars the iron abundance is being
underestimated.
Note that the \temp\ upper limit for \Space\ is 7400~K, which means that
for any spectrum of a higher temperature star \Space\ is expected to exit
with no result. Moreover, we must remember that the \Space\ line list was
built using 5 standard stars, the hottest of which is Procyon with
\temp=6554~K. As a consequence for \temp\ higher than that of Procyon the
stellar spectra may contain absorption lines that are not included in
the line list, which could potentially lead to systematic errors.\\

For the sake of completeness, we also mention a S/N dependent bias
against low metallicity stars. The lower the metallicity of a star,
the weaker are its absorption lines.  This means that for a star of lower
metallicity, the S/N must be higher in order to make the lines identifiable
through the noise. If too many lines are not identifiable, the analysis can
fail, leading to an underestimation of the number of low metallicity stars.
To explain the bias better, in the following we
propose a different point of view of this problem.\\
We usually refer to the S/N
as the ratio between the continuum level and the noise level.  However, in
measuring the strength\footnote{Here the term ``strength" refers to
the intensity of the absorption line. This can be also taken to mean
as the ratio between
the depth of the line over the continuum level and a proxy of the EW.} 
of an absorption line, we should consider
the ratio between the line strength and the noise level, a ratio that we may
call ``line strength-to-noise" ratio (L/N).  For a fixed level of noise, the L/N
diminishes with the line strengths.  Therefore, for small L/N the
measured strength of the line is more uncertain.  If L/N$<$3 we
can no longer firmly detect the line. \Space\ selects the lines (to be used
to build the spectrum model)
as a function of their detectability: if a line (or a collection of blended
lines) has an expected maximum strength much
smaller than the noise, then it is neglected. It follows that the noise level
sets a lower limit to the strength of the detectable
absorption lines and those which lie under this limit are not
considered anymore. Because the strengths
of the lines diminish with the metallicity of the stars, the number of
the absorption lines considered diminishes with the metallicity of the star.
Eventually, the number of measurable lines becomes too small
to derive the stellar parameters and the analysis fails.
This causes a bias against low
metallicity stars, and this bias is greater for lower S/N
spectra. The same bias has a greater effect on hot dwarfs, as opposed
to cold giants, because the absorption lines of the former have lower
EWs (for a given metallicity).
Computing a reliable correction for this bias would require complex
modeling and hence we do not tackle the problem here.

\subsection{Comparison between \Space\ and LASP} \label{sec:comparison_lasp}

\begin{figure*}[h]
\begin{minipage}[h]{18cm}
\centering
\includegraphics[bb=91 288 358 443,width=12cm]{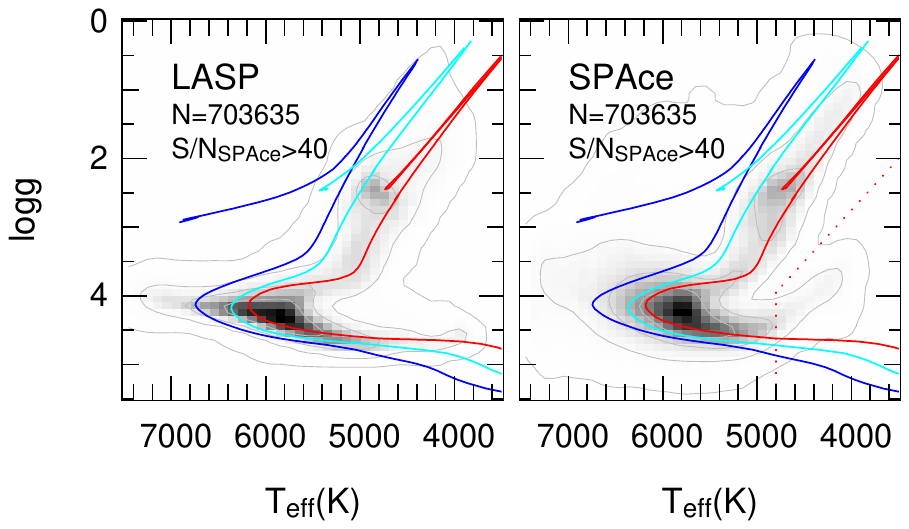}
\end{minipage}
\begin{minipage}[h]{18cm}
\centering
\includegraphics[bb=91 288 358 443,width=12cm]{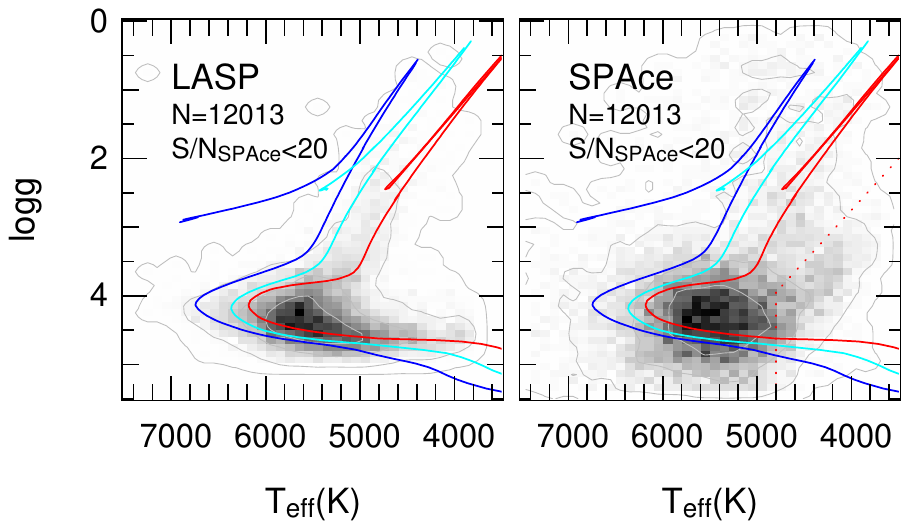}
\caption{Density distribution of the LAMOST spectra in the
(\temp,\logg) plane for LASP (left panel) and \Space\ (right panel)
for stars with \SNspace$>$40 (top) and \SNspace$<$20 (bottom).
The red, light blue, and blue lines represent the same isochrones as
in Fig.~\ref{TG_apo_space_calib}.
The red dotted line highlights the region where the stars have
systematically underestimated gravities. This is discussed in
Section~\ref{sec:systematics}.}
\label{TG_LASP_space}
\end{minipage}
\begin{minipage}[h]{18cm}
\centering
\includegraphics[bb=76 283 532 528,width=12cm]{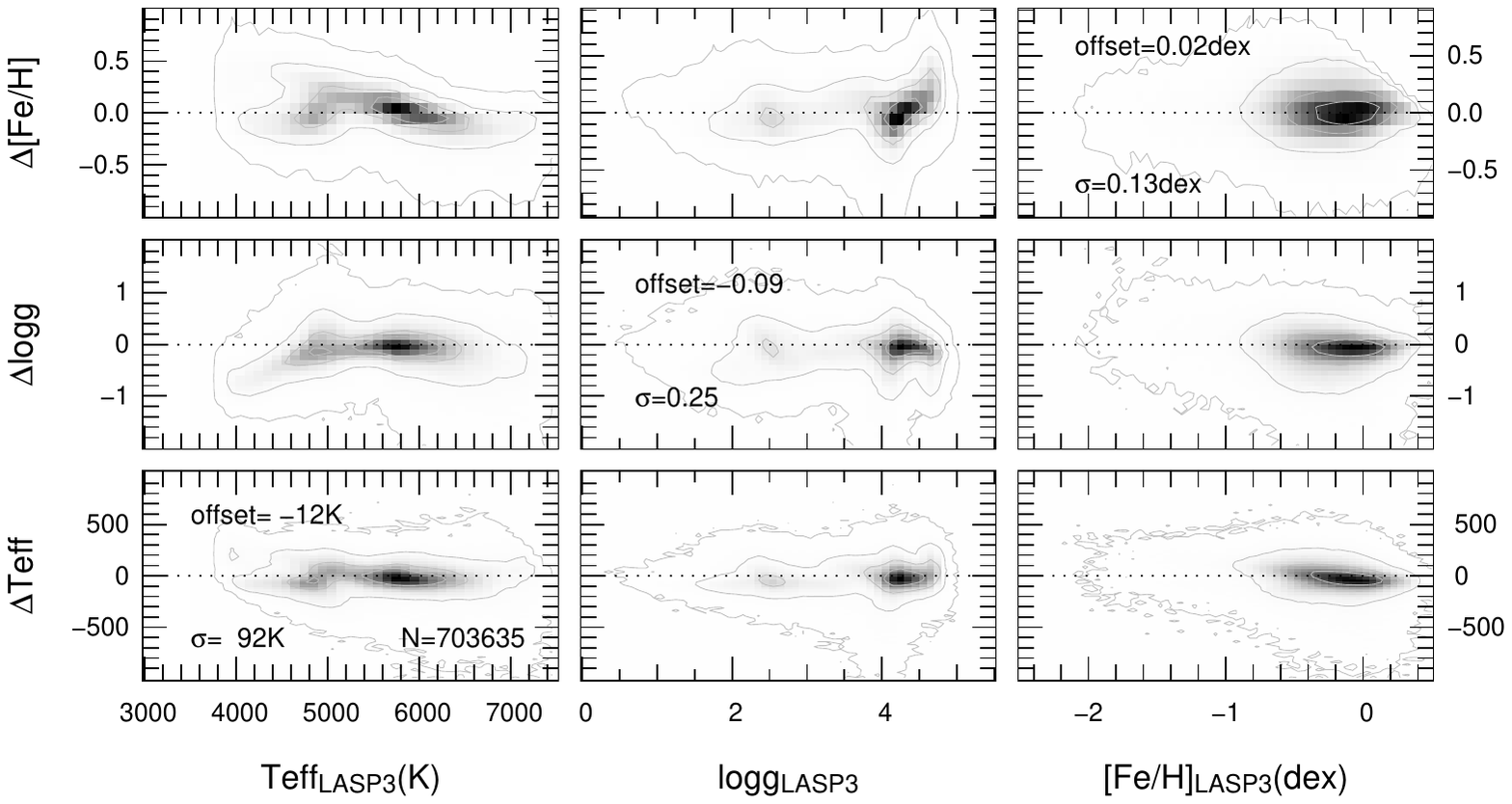}
\caption{Residuals (\Space\ minus LASP) of the stellar parameters with
respect to the LASP values for stars with \SNspace$>$40.}
\label{corr_LASP_space_SN40}
\end{minipage}
\end{figure*}

We now compare the performance of \Space\ and the official LAMOST
pipeline LASP\footnote{Here we adopt the parameters derived using the DR3
version of LASP.} (Luo et al. \citealp{luo}), using two different methods
applied to the same spectra. The first issue to consider is the total
number of converged spectra. In this aspect the performance of the two
pipelines is similar. The total number of converged spectra is
1\,125\,722 for LASP and 1\,097\,010 for \Space. The ability of each
pipeline to process low-S/N spectra is also similar; if we consider
the 506\,658 stars with LAMOST SNRI of between 20 and 40, we find that
LASP converges for 297\,597 while \Space\ converges for 301\,939. Note
that not all of these spectra will have unusable parameters; of these 
301\,939 stars, 70\% have \SNspace$>$30 and 38\% have \SNspace$>$40.
On the top of Fig.~\ref{TG_LASP_space} we compare the LASP
and \Space\ distributions in the (\temp, \logg) plane for all DR1
spectra that have \SNspace$>$40 and parameters estimated by both
codes. There are some differences, such as i)
for cool dwarf stars \Space\ derives \logg\ values that are
systematically too low (this is clearly detectable by eye at \temp$<$4800K,
but it extends, at a lesser degree, to higher \temp), ii) for
\temp$\gtrsim5600$~K the LASP gravity distribution unnaturally peaks
at \logg$\sim$4.2, iii) most of the LASP red giant branch stars do not
have \logg\ values lower than $\sim$2 while the \Space\ gravities
extend to lower values. Besides, the parameters from LASP closely
follow the isochrones, even for very low S/N where we would expect the
errors to be larger (see bottom panels of Fig.~\ref{TG_LASP_space}). 
This is due to one of the LASP features, i.e., stars are 
placed on areas of the parameter space where the
standard stars (which are employed as templates) lie. If there are no
templates for a particular location in parameter space
(for instance on the horizontal branch or far from the isochrones),
then the LASP pipeline is unable to place a star there.
As a consequence the LASP pipeline renders a more good-looking
distribution, but it can also create hidden systematic errors for
those stars that are not represented in the template sample. 
For low S/N spectra the LASP parameters lack the natural dispersion
expected for such spectra.
Fig.~\ref{corr_LASP_space_SN40} shows a detailed comparison of
the two approaches. They agree fairly well for \temp\ and \logg\, although
for cool dwarfs we find the systematic errors already
discussed in Section~\ref{sec:systematics}. The good general agreement between
the two [Fe/H] values is shown in the same figure, together with the known
systematic offsets for both extremes of the temperature range for dwarf
stars (over- and under-estimated for \temp$<$5500 and \temp$>$6500~K,
respectively).

\begin{figure*}[h]
\begin{minipage}[h]{18cm}
\centering
\includegraphics[bb=91 288 358 443,width=12cm]{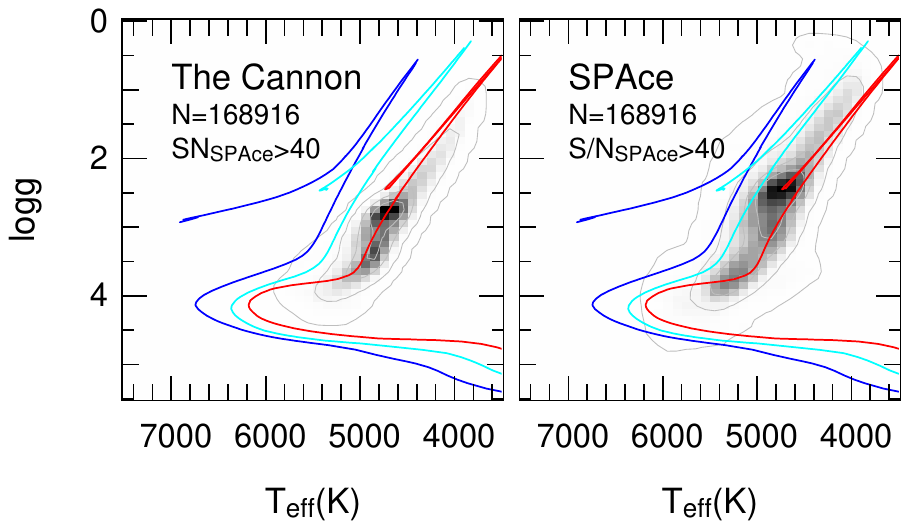}
\end{minipage}
\begin{minipage}[h]{18cm}
\centering
\includegraphics[bb=91 288 358 443,width=12cm]{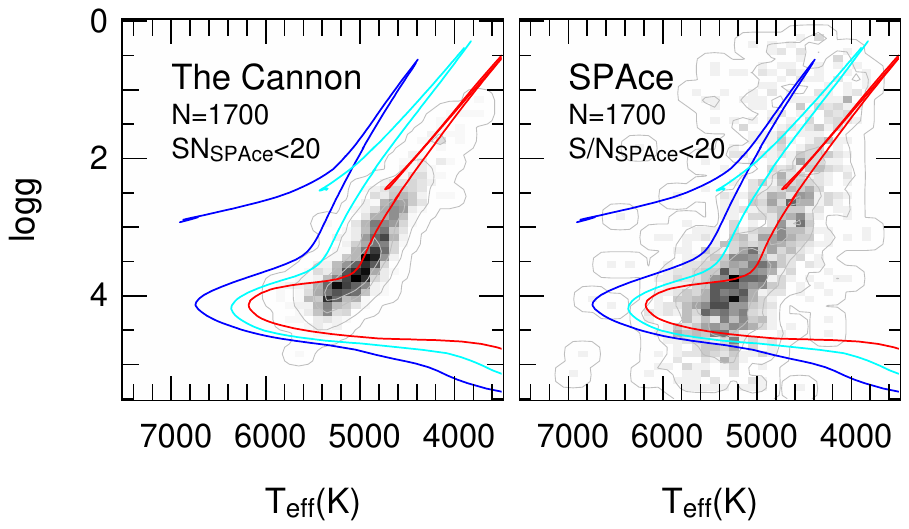}
\end{minipage}
\caption{Density distribution of the LAMOST spectra in the
(\temp,\logg) plane for {\it The Cannon} (left panel) and \Space\ (right panel)
for stars with \SNspace$>$40 (top) and \SNspace$<$20 (bottom). 
The red, light blue, and blue lines represent the same isochrones as in
Fig.~\ref{TG_apo_space_calib}.}
\label{TG_cannon_space}
\end{figure*}

\begin{figure*}[h]
\centering
\includegraphics[bb=5 283 582 593,width=13cm]{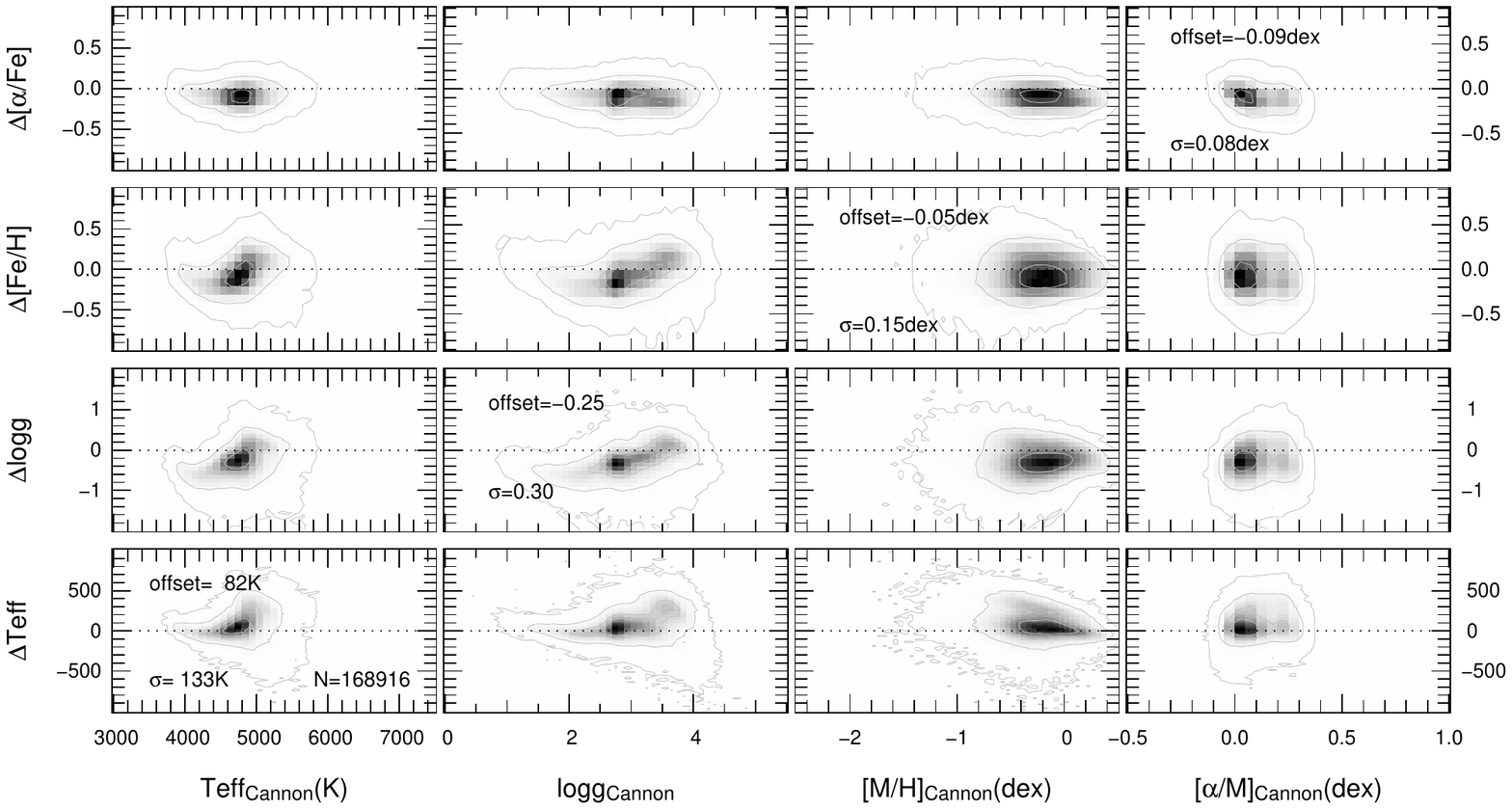}
\caption{Residuals (\Space\ minus {\it The Cannon}) of the stellar parameters with
respect to the {\it The Cannon} values for stars with \SNspace$>$40.}
\label{corr_cannon_space_SN40}
\end{figure*}

\subsection{Comparison between \Space\ and {\it The Cannon}}
\label{sec:comparison_cannon}

Recently Ho et al. (\citealp{ho2}) used {\it The Cannon}
(Ness et al., \citealp{ness}) to derive stellar parameters and abundances of
the alpha-elements and the individual elements C and N of giant stars from
the LAMOST DR2 internal data release.  {\it The Cannon} has been trained on
a set of spectra of objects that LAMOST has in common with APOGEE.  In their
work, Ho et al. demonstrated the consistency between their stellar
parameters and those obtained by APOGEE from high-resolution spectra.
As previously done with the LASP pipeline, we now compare the Ho et
al.  results with the ones obtained with \Space.  Out of the 454\,450
spectra considered by Ho et al., we have in common 219\,360 spectra
belonging to LAMOST DR1.

The distribution in the (\temp, \logg) plane
of the spectra with \SNspace$>40$ is shown on the top of
Fig.~\ref{TG_cannon_space}.
{\it The Cannon} places RC stars in a similar position to APOGEE
(Fig.~\ref{TG_apo_space_calib}), although the former are located at
slightly higher gravity, in a position more suitable for RGB stars.
We investigate this further using the asteroseismic \logg\ values from
Huber et al. \citep{huber} for the two samples of RC and RGB stars classified
by Stello et al. (which were introduced in Section~\ref{sec:kepler}, 
see Fig. \ref{fig:stello}).
From its high-resolution spectroscopy APOGEE finds offsets of 0.14 and
-0.03 dex for the RC and RGB, respectively (from samples of 1045 and
661 stars, respectively). These results are to be expected as APOGEE
gravities are calibrated onto the asteroseismic RGB sample of Pinsonneault et
al. \citep{pinsonneault}, hence the good agreement for the RGB and
systematic overestimation for the RC (see Fig. 4 of Holtzman et
al. \citealp{holtzman}).
If we now consider the low-resolution LAMOST spectra as analyzed by 
{\it The Cannon}, we find that its performance is subject to larger
systematic errors with offsets of 0.31 and 0.15 dex, respectively
(from samples of 265 and 112 stars, respectively). This can be
compared to the performance of \Space\ (Fig.~\ref{fig:stello}), which
has offsets of only -0.01 and -0.08 dex, respectively (from 
1366 and 928 stars, respectively). The scatter for \Space\ is
marginally larger than for {\it The Cannon} (0.13 vs 0.10 dex), but
the systematic errors are significantly less. The fact that the
systematic errors are smaller than even the high-resolution data from
APOGEE demonstrates the efficacy of \Space\ for recovering accurate
gravities from LAMOST spectra. Note that other machine learning methods
have had more success at fitting gravities (e.g. Liu et al. \citealp{liu}), but a
comprehensive comparison of all pipelines is beyond the scope of this
study.

The correlations between the residuals \Space\ minus
{\it The Cannon} and the {\it The Cannon}
parameters, as shown in Fig.~\ref{corr_cannon_space_SN40}, are similar to
those obtained for the APOGEE data (Fig.~\ref{corr_apo_SN40}). This is
not surprising since {\it The Cannon} has been trained with a set of
spectra labeled with the APOGEE data. The same systematics are
observed in gravity, [Fe/H], and \logg. However, there is a group of
stars for  which \Space\ assigns a higher \temp\ than {\it The Cannon}
(stars at {\it The Cannon} \logg$\sim3.5$) and the same stars also have
higher \Space\ \logg. {\it The Cannon} locates these stars on the
red giant branch while \Space\ classifies them as sub-giants.
Unfortunately, in this region of parameter space there are too few
stars with asteroseismic gravities to clarify the origin of this
offset.

On the bottom of Fig.~\ref{TG_cannon_space} we report the distributions of low \SNspace\
stars in the (\temp, \logg) plane for {\it The Cannon} and \Space. 
While at low S/N the \Space\ parameters show a scatter around the
isochrones (as expected) the lack of dispersion shown by {\it The
Cannon} is similar to the LASP pipeline. We make the same comment as
before, namely that codes relying on training sets of spectra
that do not uniformly cover the parameter space tend to assign stellar
parameters to regions covered by the training sets. This means that
they lack the natural scatter due to the uncertainties coming from
working at low S/N. This aspect should be further investigated to
establish the presence (or lack) of possible systematic errors as a
function of the S/N, particularly in regions of parameter space that
are not covered by stars in the training set.

\section{Conclusions}\label{sec:conclusions}

We have used the code \Space\ to derive stellar parameters for 1\,097\,231
stellar objects from the LAMOST DR1 catalog.  In addition to the parameters
\temp, \logg, and [Fe/H] (which are also given by the LAMOST LASP pipeline)
we have also derived the alpha abundance [$\alpha$/H].  By
comparing our results to high precision parameters for stars from surveys
such as APOGEE, the Gaia-ESO survey, the Geneva-Copenhagen Survey, and
the Kepler mission, we have confirmed the ability of \Space\ to derive stellar
parameters and chemical abundances for FGK stars.  This has also allowed us
to demonstrate the robustness of our results.  We have highlighted the
presence of some systematic errors in our results, such as the
overestimation of \logg\ for cold dwarfs, an underestimation of gravity for
\logg$\lesssim 2$, and a bias against hot metal rich dwarfs.  We have also
shown that the \Space\ error estimates look reliable when compared to the
residuals between our results and these reference parameters.

We compared our results to other pipelines which have been used on
LAMOST spectra, namely the LASP and {\it The Cannon} pipelines. The
comparison between the three pipelines can be summarized as follows:\\
\begin{itemize}
\item  \Space\ shows a systematic error in \logg\ for dwarf stars
  cooler than \temp$<4800$~K while LASP and {\it The Cannon} perform
  well in this parameter region.
\item LASP shows an excess of stars
  with \logg$\sim4$ that is not expected and not seen in
  the \Space\ results.
\item The LASP results are biased against giant stars with \logg$<2$,
  which is unexpected as these stars are detected by \Space\ and {\it
    The Cannon}.
\item The position of the RC stars is in good agreement with the
  isochrones for LASP and \Space\, while it is overestimated in \logg\
  by {\it The Cannon} (very likely due to the APOGEE training set used
  by {\it The Cannon}, which shows the same systematic).
\item At low S/N, \Space 's stellar parameters show the natural
  dispersion expected from spectra that hold little or no
  information. The other two pipelines follow the isochrones much more
  closely, which is a consequence of how they estimate parameters 
  (i.e. both techniques use of training sets whose distribution closely match the
  isochrones) . This
  artificial lack of dispersion is worrying and the results from these
  pipelines require further investigation, in order to check whether
  this behavior introduces any systematic errors.\\
\end{itemize}

The catalog presented here is publicly available at the
LAMOST\footnote{http://dr1.lamost.org/doc/vac} and
CDS\footnote{http://cds.u-strasbg.fr} websites.
In future we would like to extend the catalog to more recent (and
larger) LAMOST data releases, using a new version of \Space\ designed
to overcome the limitations identified in this current study.

\acknowledgments

This work was supported by the National Key Basic Research Program of
China ``973 Program'' 2014 CB845700 \& CB845702, and by the National
Natural Science Foundation of China (NSFC) under grants 11673083 (PI:
M.C. Smith), 11333003 (PI: Shude Mao) \& 11373054 (PI: L. Chen).  

C.B. and E.K.G. were supported by Sonderforschungsbereich SFB 881 ``The Milky
Way System" (sub-project A5) of the German Research Foundation (DFG).
D.S is the recipient of an Australian Research Council Future
Fellowship (project number FT1400147).

Guoshoujing Telescope (the Large Sky Area Multi-Object Fiber
Spectroscopic Telescope, LAMOST) is a National Major Scientific
Project built by the Chinese Academy of Sciences. Funding for the
project has been provided by the National Development and Reform
Commission. LAMOST is operated and managed by the National
Astronomical Observatories, Chinese Academy of Sciences.

\appendix
\section{Identified problems in \Space}\label{appendix:bugs}

During the progress of this work we found one bug in the LAMOST version of
\Space\ that affects the mechanism with which \Space\ aborts the analysis of
spectra with large RV.  This involves the offset limit in radial
velocity correction beyond which \Space\ stops, and affects the LAMOST version
only (the official public version of \Space\ does not have this bug).
Since the LAMOST version of \Space\ handles both the blue and the red part of the
spectrum, \Space\ stops when one or both parts show a wavelength offset
(due to the RV) larger than 1.27FWHM. For this comparison, the source code
translates the RV into \AA\ at the wavelength of 5500\AA\ for both blue
and red parts. This is a mistake because in this way the blue and red RV limits 
at which \Space\ stops are different, with the blue part giving the
more stringent limit (which is $\sim \pm200$\kmsec\ for an average
FWHMb$\sim$2.9\AA).
However it must be stressed that, even if this bug were not present, 
the blue and red RV limits can never be constant because the limit depends
on the FWHMb and FWHMr, which are not equal within one spectrum, nor 
among different spectra. This does not affect the robustness of any of the
parameters derived by \Space\ but it can create a bias against stars with
high radial velocity (with respect to the LSR). 
This will be corrected in our next version of LAMOST \Space.

\section{Validation plots}\label{appendix:plots}
For the sake of completeness, here we include additional
plots illustrating the comparison between \Space\ and the reference
data sets described in Section~\ref{sec:validation}.
These plots show the detailed correlations between parameters and are
helpful in identifying potential problems with \Space\ parameters.

\begin{figure*}[h]
\centering
\includegraphics[bb=0 282 582 600,width=16cm]{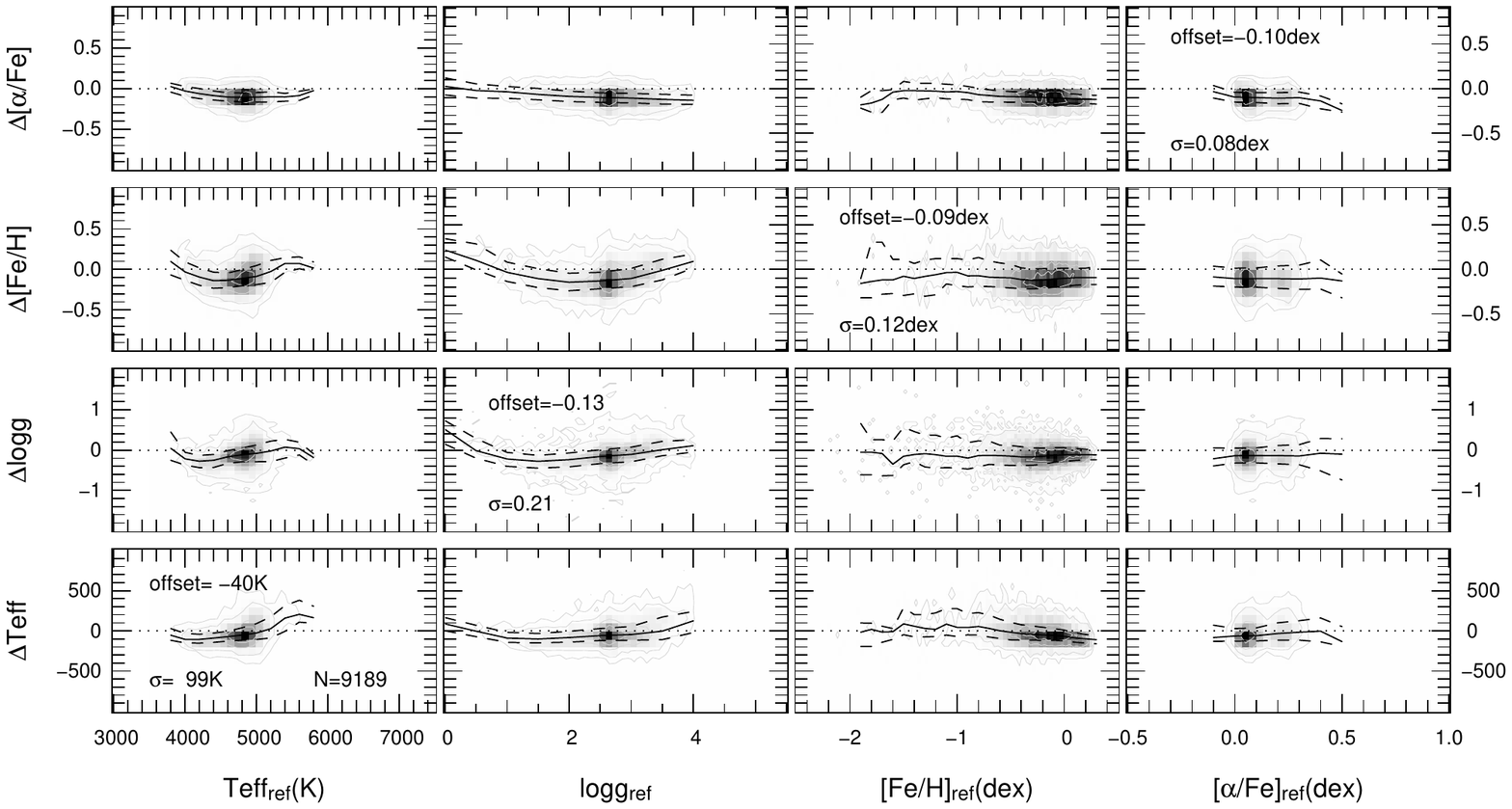}
\caption{Residuals (\Space\ minus ASPCAP/APOGEE) of the stellar parameters with
  respect to the reference parameters for stars with \SNspace$>$40. The
solid black lines show the median of the residuals as a function of
the reference stellar parameter, while the upper and lower
dashed lines denote the interval holding 68\% of the sample. These
have been constructed using bins of width 200~K in \temp, 0.5 in
\logg, 0.1~dex in \met\ and \aFe.}
\label{corr_apo_SN40}
\end{figure*}

\begin{figure*}[h]
\centering
\includegraphics[bb=94 293 522 746,width=16cm]{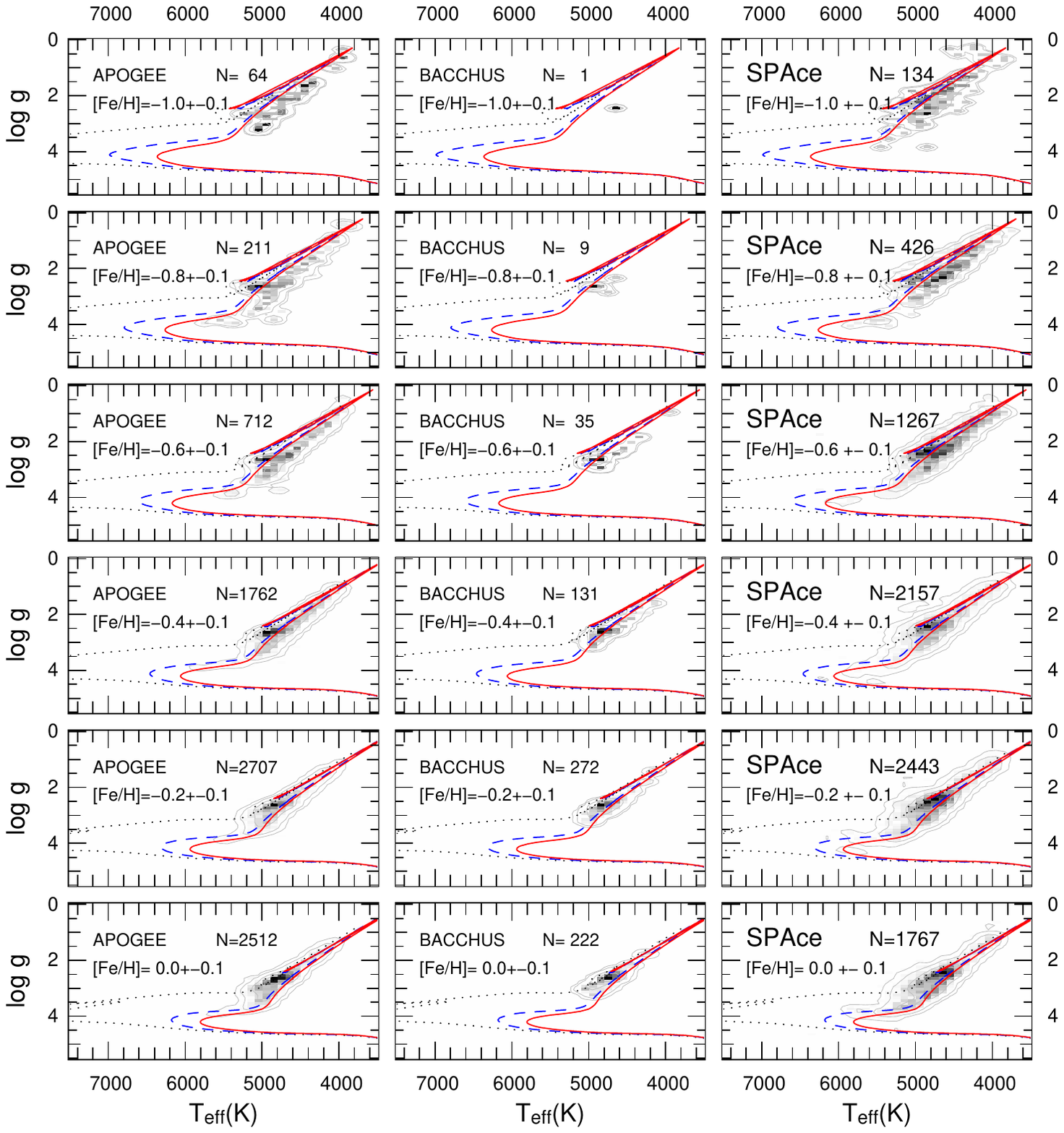}
\caption{Distribution of stars in the (\temp,\logg)-plane for the
APOGEE/ASPCAP, BACCHUS, and \Space\ samples, divided into bins of [Fe/H]
with half-width of 0.1~dex. The average [Fe/H] of the bins increases
from top to bottom. The dotted black, blue dashed, and solid red lines
are Bressan et al. isochrones of 1, 5 and 10~Gyr, respectively, with
metallicity corresponding to the [Fe/H] bin. The stars included here
have \SNspace$>$40. The grey-scale indicates the relative
density of stars and the contours enclose 34, 68, 95, and 99\%
of the sample.}
\label{plot_apo_bacchus_met_bin}
\end{figure*}

\begin{figure*}[h]
\begin{minipage}[h]{6cm}
\includegraphics[width=\hsize]{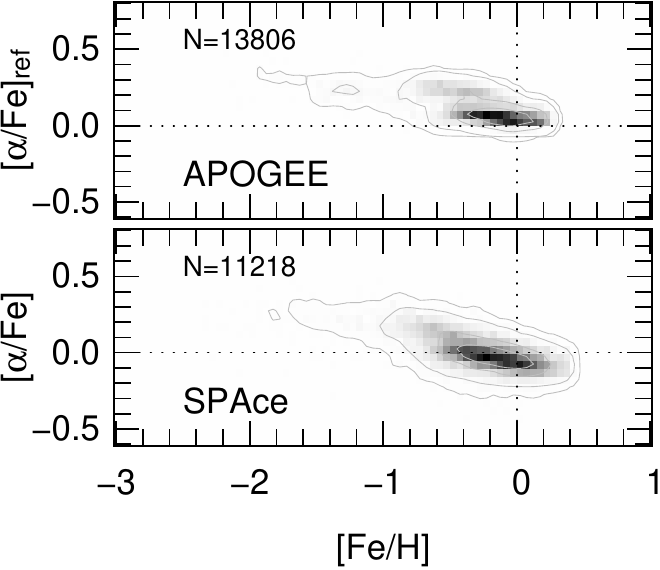}
\caption{$\alpha$/Fe distributions for APOGEE stars with spectra also
  observed by LAMOST (and for which \SNspace$>$40). The top panel shows the ASPCAP parameters from
  the APOGEE spectra, while the bottom panel shows
  the \Space\ parameters from the LAMOST spectra.}
\label{apo_space_aFe}
\end{minipage}
\begin{minipage}[h]{12cm}
\includegraphics[bb=83 287 610 555,width=\hsize]{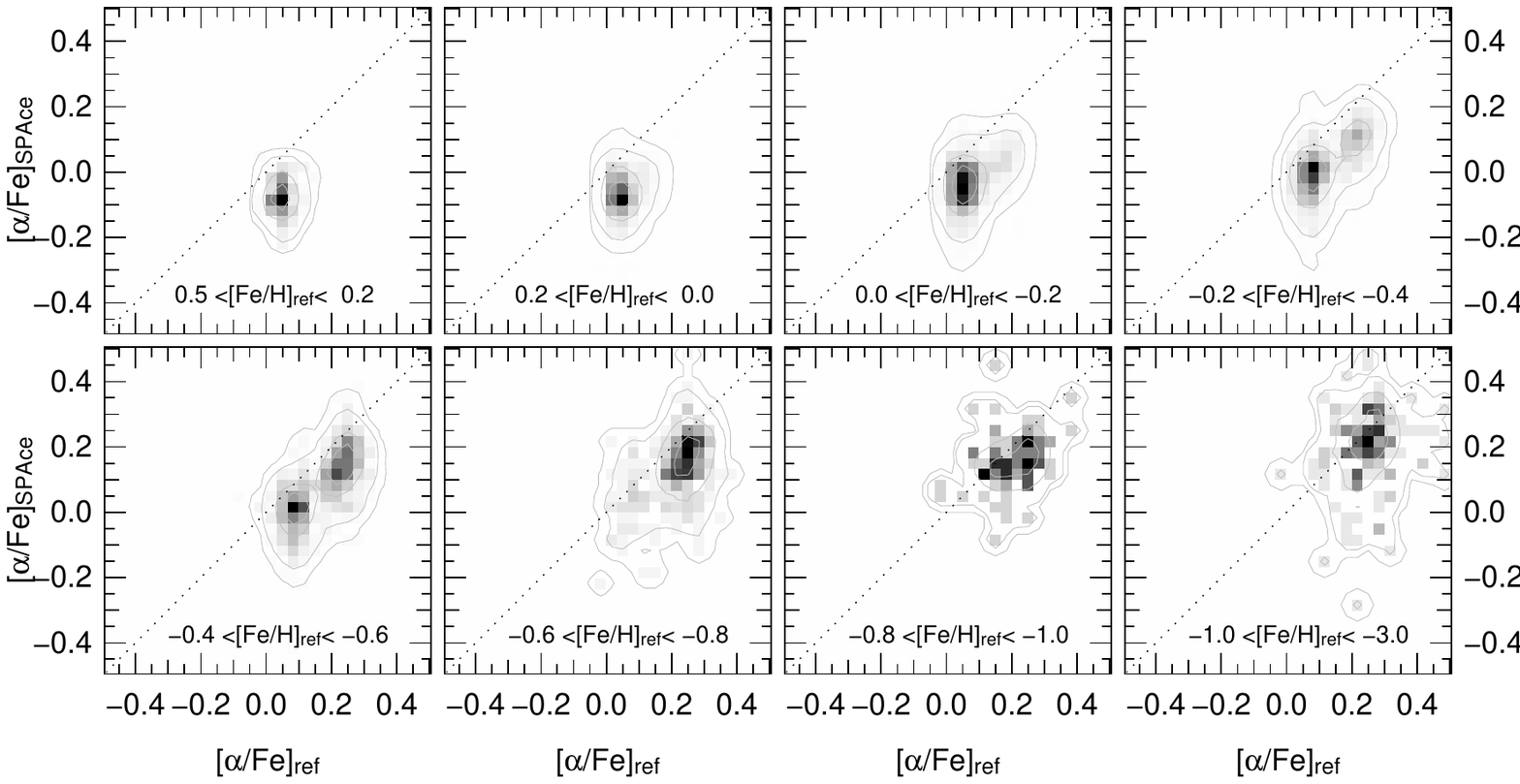}
\caption{Comparison between $\alpha$/Fe derived by ASPCAP and \Space,
divided in ASPCAP [Fe/H] bins, for stars with \SNspace$>$40.}
\label{comp_alphaFe_apo}
\end{minipage}
\end{figure*}

\begin{figure*}[h]
\centering
\includegraphics[bb=0 282 433 530,width=16cm]{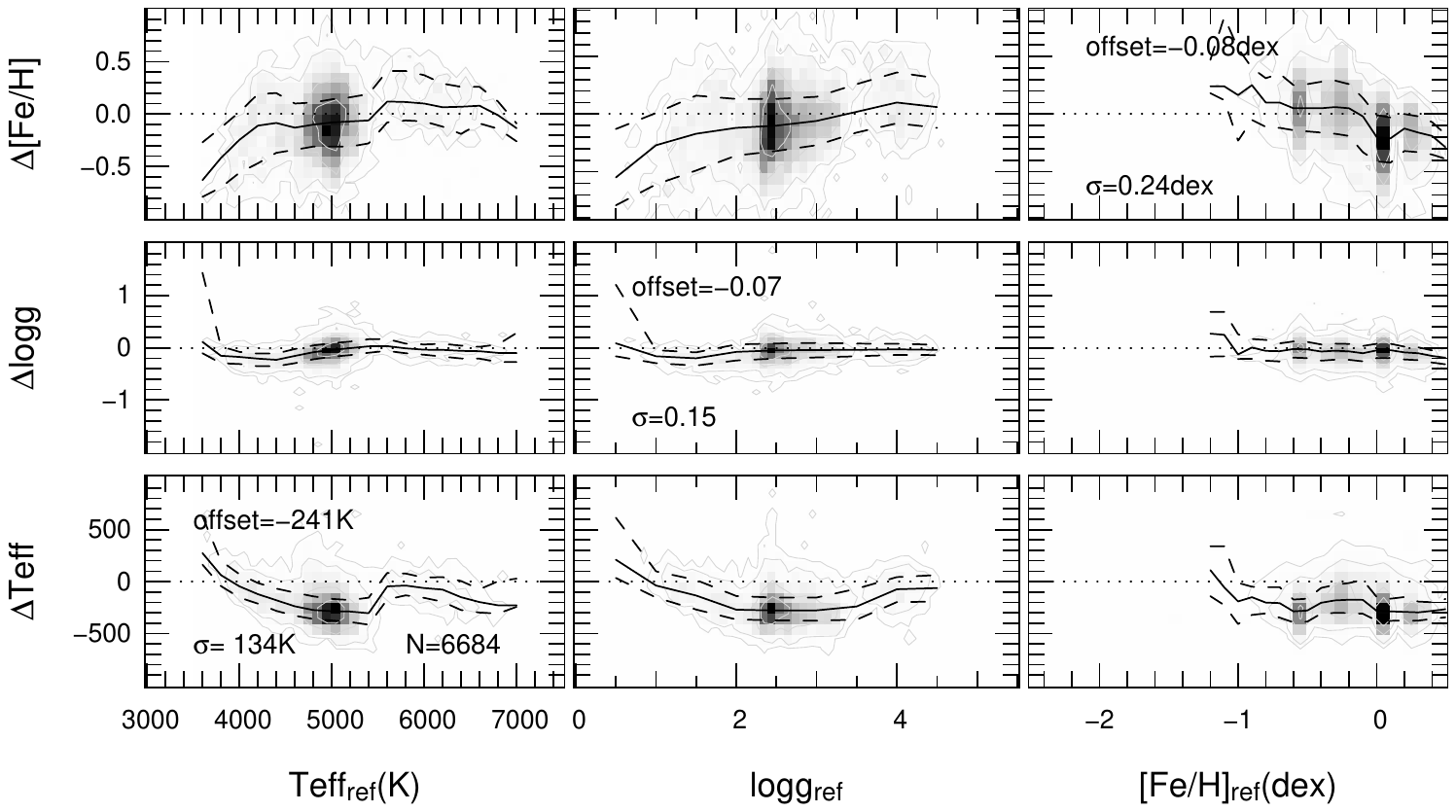}
\caption{Residuals (\Space\ minus the Kepler stars of Huber et al.) of the stellar parameters with
respect to the reference parameters for stars with \SNspace$>$40. Black
and dashed lines as in Fig.~\ref{corr_apo_SN40}.}
\label{corr_huber_SN40}
\end{figure*}

\begin{figure*}[h]
\centering
\includegraphics[bb=0 282 584 597,width=16cm]{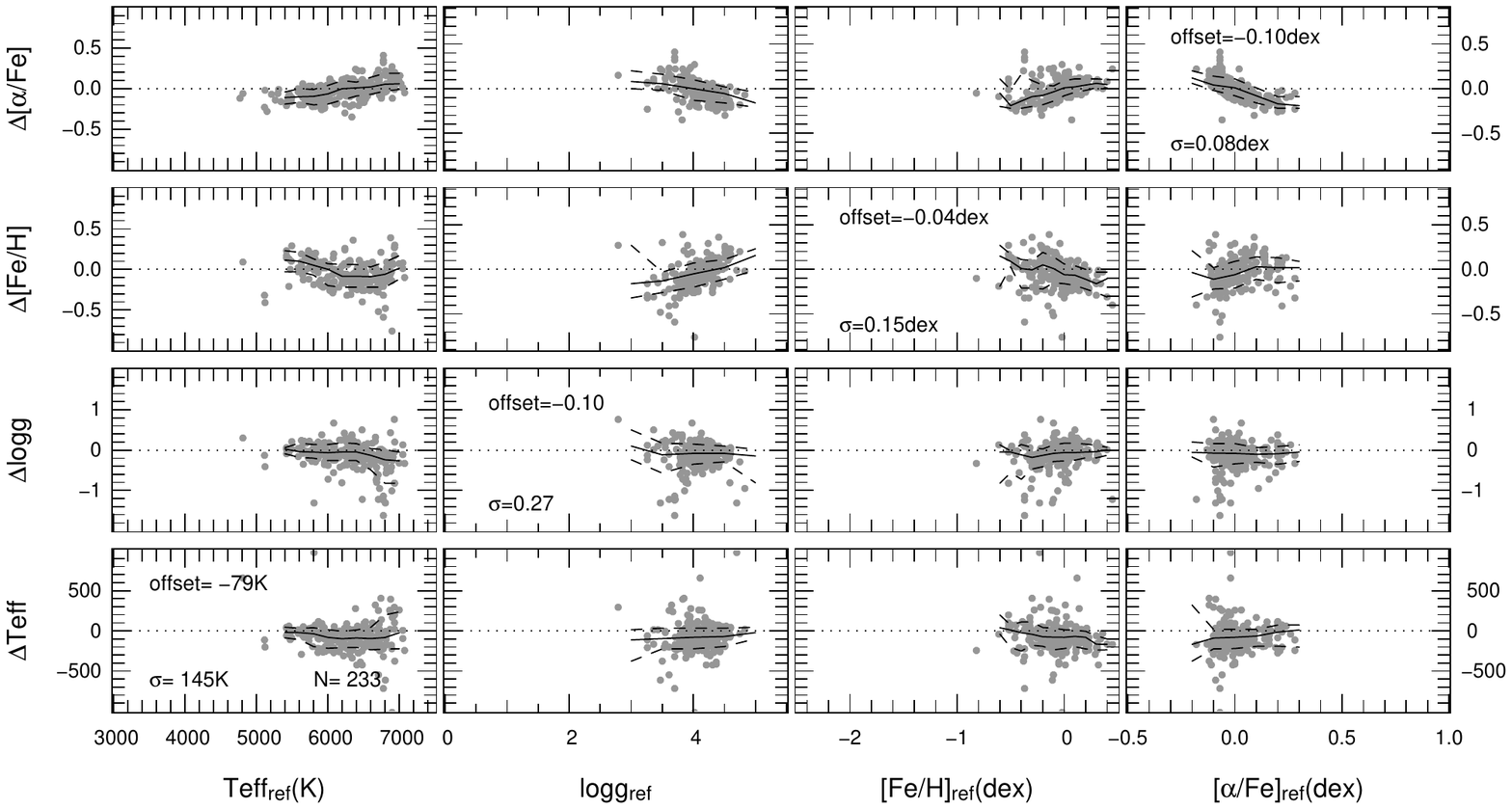}
\caption{Residuals (\Space\ minus Casagrande et al.) of the stellar parameters with
respect to the reference parameters for stars with \SNspace$>$40. Black
and dashed lines as in Fig.~\ref{corr_apo_SN40}.}
\label{corr_GCS_SN40}
\end{figure*}

\begin{figure*}[h]
\centering
\includegraphics[bb=0 282 582 600,width=16cm]{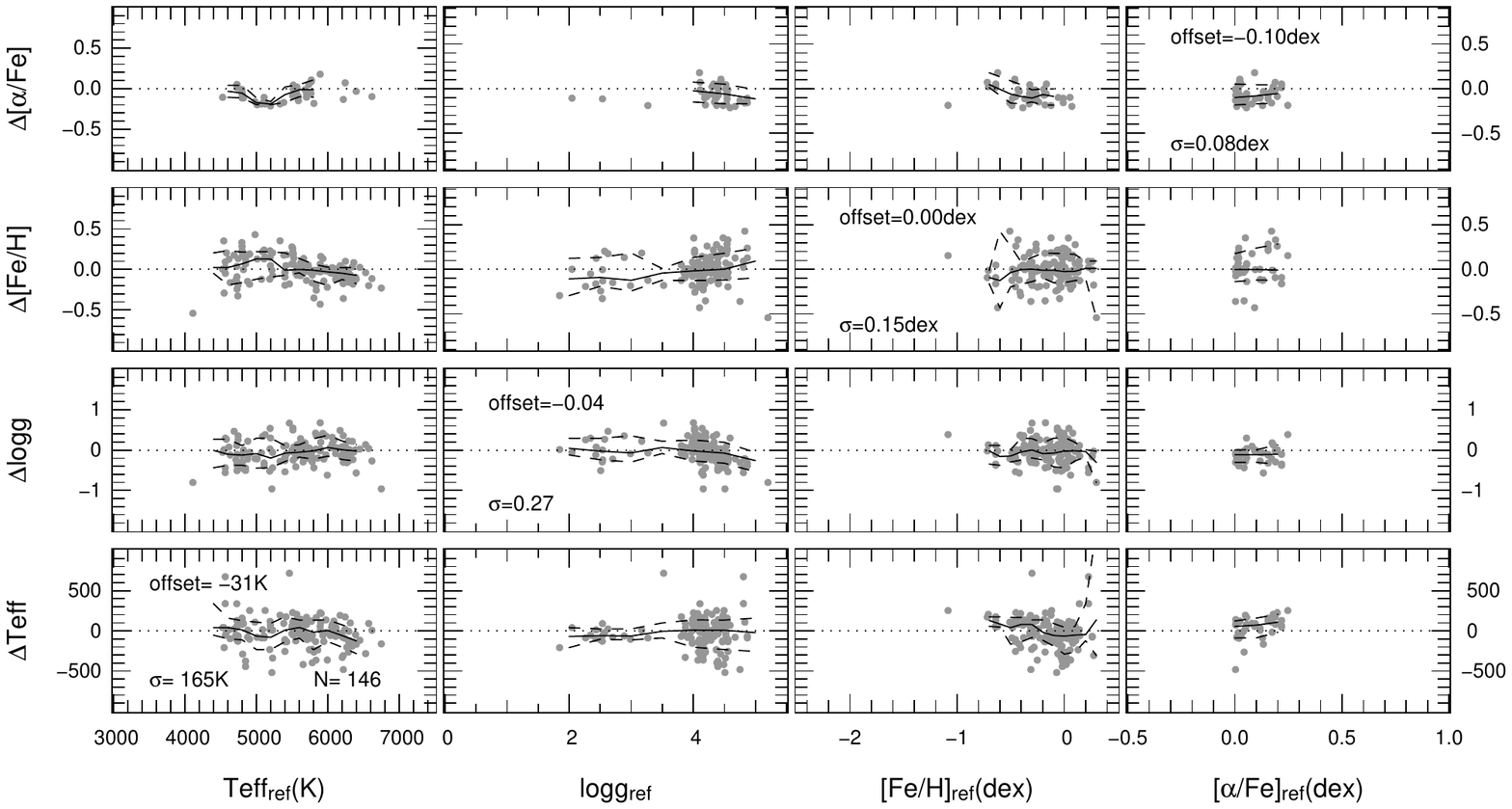}
\caption{Residuals (\Space\ minus GES) of the stellar parameters with
respect to the reference parameters for stars with \SNspace$>$40. Black
and dashed lines as in Fig.~\ref{corr_apo_SN40}.}
\label{corr_GES_SN40}
\end{figure*}

\begin{figure*}[t]
\begin{minipage}[h]{17cm}
\centering
\includegraphics[bb=39 428 573 555,width=\hsize]{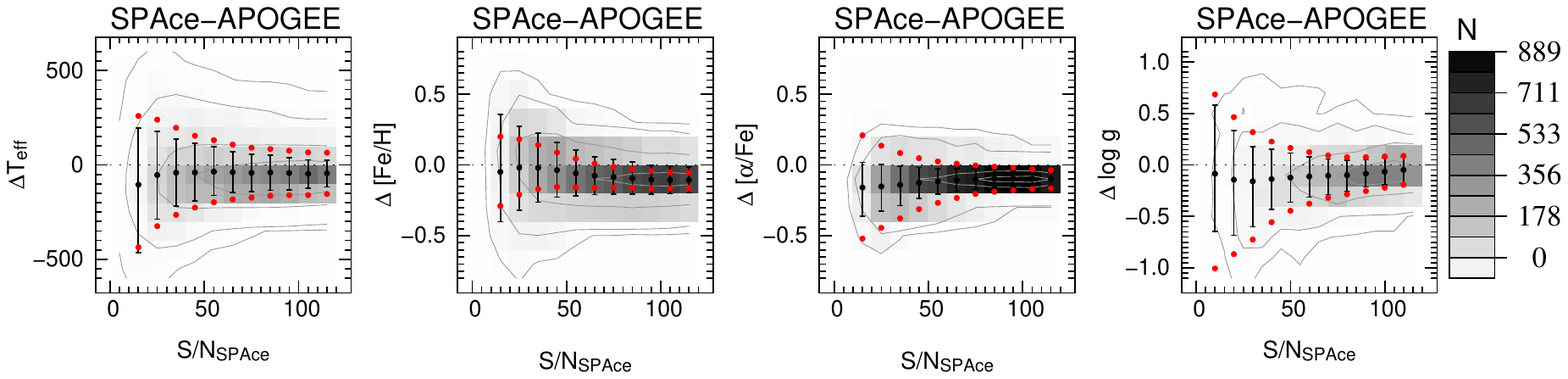}
\end{minipage}
\begin{minipage}[h]{17cm}
\centering
\includegraphics[bb=39 428 573 555,width=\hsize]{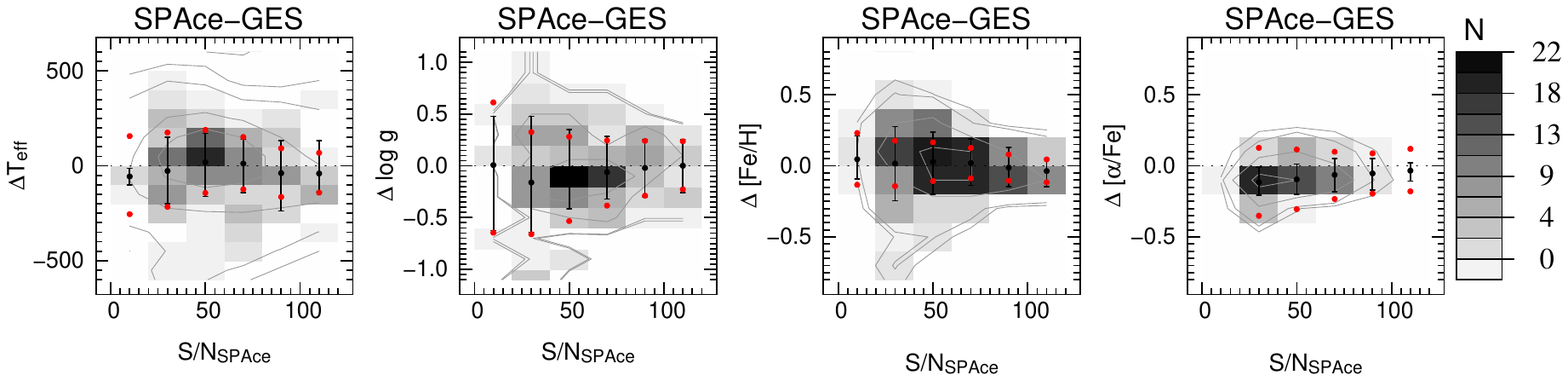}
\end{minipage}
\caption{{\bf Top:} Density distributions for the discrepancies (\Space\ minus reference)
for data sets dominated by giant stars. The left and middle panels show the
comparison of \Space\ and ASPCAP/APOGEE, while the right panel
shows \Space\ compared with the Kepler stars of Huber et al.
The dots with error bars represent the average
discrepancy and 1$\sigma$ of the distribution for each bin in \SNspace, while
the red dots represent the average uncertainties ($\pm1\sigma$)
computed by adding the \Space\ and reference errors in
quadrature. The grey-scale indicates the density of points per
  pixel (see bar on right), while the contours enclose 34, 68, 95, and 99\% of the sample
and all stars have \SNspace$>$40. {\bf Bottom: } 
Density distributions for the discrepancies between \Space\ and GES, i.e., 
for a sample dominated by main-sequence stars. 
Symbols, grey-scale bar, and contours are as in the top panel.}
\label{discr_SN_errors_giants}
\end{figure*}

\clearpage


\end{document}